\documentclass{aa}

\usepackage[english]{babel}
\usepackage{graphicx}
\usepackage{amsmath}
\usepackage{amssymb}
\usepackage{txfonts}
\usepackage{wasysym}
\usepackage{gensymb}

\usepackage[colorlinks=true, linkcolor=blue, citecolor=blue, urlcolor=blue]{hyperref}
\usepackage[singlelinecheck=true,format=default,font=small,labelfont=bf]{caption}
\usepackage{natbib}
\usepackage{color}
\usepackage{natbib}
\usepackage{multirow}
\usepackage{array}

\bibpunct{(}{)}{;}{a}{}{,} 

\begin{document}

\title{NGC4370: a case study for testing our ability to infer dust distribution and mass in nearby galaxies}

\author{S. Viaene\inst{1} \and G. De Geyter \inst{1} \and M. Baes \inst{1} \and J. Fritz \inst{1,2} \and G.J. Bendo \inst{3} \and M. Boquien\inst{4} \and A. Boselli\inst{5}  \and S. Bianchi \inst{6} \and L. Cortese\inst{7} \and P. Côté \inst{8} \and J.-C. Cuillandre \inst{9} \and I. De Looze\inst{1,10} \and S. di Serego Alighieri \inst{6} \and L. Ferrarese \inst{8} \and S. D. J. Gwyn \inst{11} \and T. M. Hughes\inst{1} \and C. Pappalardo \inst{12}}

\institute{Sterrenkundig Observatorium, Universiteit Gent, Krijgslaan 281, B-9000 Gent, Belgium\\
        \email{sebastien.viaene@ugent.be}
        \and
        Centro de Radioastronom\'\i a y Astrof\'\i sica, CRyA, UNAM, Campus Morelia, A.P. 3-72, C.P. 58089, Michoac\'an, Mexico
        \and
        Jodrell Bank Centre for Astrophysics, School of Physics and Astronomy, University of Manchester, Oxford Road, Manchester M13 9PL, UK
        \and
        Institute of Astronomy, University of Cambridge, Madingley Road, Cambridge, CB3 0HA, UK
        \and
        Laboratoire d'Astrophysique de Marseille, UMR 6110 CNRS, 38 rue F. Joliot-Curie, F-13388 Marseille, France
        \and
        Osservatorio Astrofisico di Arcetri – INAF, Largo E. Fermi 5, 50125 Firenze, Italy
        \and
        Centre for Astrophysics \& Supercomputing, Swinburne University of Technology, Mail H30 – PO Box 218, Hawthorn, VIC 3122, Australia 
        \and
        National Research Council of Canada, Herzberg Astronomy and Astrophysics, 5071 W. Saanich Road, Victoria, BC V9E2E7, Canada
        \and
        Canada-France-Hawaii Telescope, 65-1238 Mamalahoa Hwy, Kamuela, HI 96743 USA
        \and
        Institute of Astronomy, University of Cambridge, Madingley Road, Cambridge, CB3 0HA, UK
        \and
        Canadian Astronomy Data Centre, 5071 West Saanich Rd, Victoria BC, V9E 2E7, Canada
        \and
        Centro de Astronomia e Astrofísica da Universidade de Lisboa, Observatório Astronómico de Lisboa, Tapada da Ajuda, 1349-018 Lisbon, Portugal
}

\abstract{A segment of the early-type galaxy population hosts a prominent dust lane, often decoupled from its stellar body. Methods of quantifying the dust content of these systems based on optical imaging data usually yield dust masses that are an order of magnitude lower than dust masses derived from the observed far IR (FIR) emission. The discrepancy is often explained by invoking a diffuse dust component that is hard to trace in the UV or optical.}{High-quality optical data from the Next Generation Virgo cluster Survey (NGVS) and FIR/sub-mm observations from the Herschel Virgo Cluster Survey (HeViCS) allow us to revisit previous methods of determining the dust content in galaxies and explore new ones. NGC 4370 is an edge-on, early-type galaxy with a conspicuous dust lane and regular morphology, making it suitable for several (semi-)analytical modelling techniques. We aim to derive the dust mass from both optical and FIR data and to investigate the need to invoke a putative diffuse dust component.}{We used different methods to determine the total dust mass in the dust lane. We used our exquisite optical data to create colour and attenuation maps, which are converted to approximate dust mass maps based on simple dust geometries. Dust masses were also derived from SED fits to FIR to sub-mm observations. Finally, inverse radiative transfer fitting was performed to investigate more complex dust geometries, such as an exponential dust disc and a dust ring and to treat the dust-starlight interaction in a self-consistent way.}{We find that the empirical methods applied to the optical data yield lower limits of $3.4 \times 10^5 M_\odot$, an order of magnitude below the total dust masses derived from SED fitting. In contrast, radiative transfer models yield dust masses that are slightly lower, but fully consistent with the FIR-derived mass. We find that the effect of a nuclear stellar disc on the derivation of the total dust mass is minor.}{Dust is more likely to be distributed in a ring around the centre of NGC 4370 as opposed to an exponential disc or a simple foreground screen. Moreover, by using inverse radiative transfer fitting, we are able to constrain most of the parameters that describe these geometries. The resulting dust masses are high enough to account for the dust observed at FIR/sub-mm wavelengths, so that no diffuse dust component needs to be invoked. We furthermore caution against interpreting dust masses and optical depths based on optical data alone, when using overly simplistic star-dust geometries and the neglect of scattering effects.}


\titlerunning{Dust in NGC 4370}
\authorrunning{S. Viaene}

\maketitle

\section{Introduction}
Early-type galaxies (ETGs) are not red and dead as was once the common understanding. Deeper observations across the electromagnetic spectrum have revealed significant amounts of matter in the interstellar medium (ISM) and even star forming activity in these objects \citep[see e.g.][]{Rampazzo2005,Combes2007,Young2011}. 
The likely mechanism to form early-type galaxies is found in major or minor mergers of spiral galaxies. During these events, interstellar matter of the progenitors is partly expelled and partly consumed in waves of fast star formation \citep[see e.g.][]{Barger1996, Delucia2006}. After a few Gyr, the galaxy relaxes to become a stable, non-star-forming object with a depleted ISM. Many ETGs, especially the more massive ones, are embedded in a halo of hot, X-ray-emitting gas \citep{Forman1985,Sarzi2013}, which is the remnant of the violent merger processes. Interstellar dust cannot survive in the harsh environment of the hot gas. Measurements of significant amounts of dust of various morphologies in over half of the ETGs \citep{Goudfrooij1994a,Goudfrooij1994b,vanDokkum1995,Ferrari1999,Tran2001,Patil2007,Finkelman2008,Kaviraj2012,Kulkarni2014} then lead to the conclusion that the dust reservoir must be replenished in some way. Since internal production mechanisms seem to fall short, a merger origin of the dust seems the most likely option \citep[see e.g. ][ for an elaborate account]{Finkelman2012}.

The dust mass determination in ETGs is complicated by the discrepancy found between the different dust mass measurements. Numerous FIR observations of ETGs  have been obtained using \textit{IRAS}, \textit{ISO,} and \textit{Spitzer} \citep[see e.g. ][]{Knapp1989,Leeuw2004,Temi2004,Temi2007,Xilouris2004}. \citet{Temi2004} found that dust masses derived from \textit{ISO} observations ($2.5 - 240~\mu$m) are more than an order of magnitude higher than \textit{IRAS}-derived ($12 - 100~\mu$m) dust masses. They argue that \textit{ISO} detects much more cold dust, which could explain the difference. Indeed, \textit{IRAS} was only sensitive to wavelengths shorter than $100~\mu$m, making it insensitive to the emission from cold ($<25$ K) dust \citep{Devereux1990}. Because cold dust is actually by far the most massive dust component in galaxies, \textit{IRAS}-derived dust masses consequently yield lower limits \citep{Gordon2010}. 

\textit{IRAS}-derived dust masses are in turn an order of magnitude higher than those computed using extinction models \citep[see e.g.][and references therein]{Patil2007, Finkelman2012}. Deriving dust masses from ultraviolet (UV), optical, and near-infrared (NIR) data can be difficult due to the effects of scattering and the complicated star-dust distribution \citep[see e.g.][]{Witt1992,Baes2001}. The estimated amount of dust is thus highly model dependent. Moreover, it requires a consistent dataset spanning a sufficient wavelength range to infer the wavelength-dependent extinction by dust. Up to now, the generally applied dust geometry was that of a dust screen in between the galaxy and the observer \citep[see e.g.][]{Goudfrooij1994b,Ferrarese2006, Patil2007, Finkelman2008, Finkelman2010}. This naturally yields lower limits of the dust mass when derived from extinction. The dust mass discrepancy with far-IR (FIR) measurements may be explained by invoking a second dust component, which follows the stellar body and is diffuse \citep[hence not detectable in the optical;][]{Goudfrooij1995,Wise1996}. 

The diffuse dust component in ETGs is not to be confused with the component of the same name found in spiral galaxies. The latter term was invoked to differentiate the dust that resides in clumps or filaments in the disks of spiral galaxies. The diffuse dust in ETGs does not refer to this low-scale height component, but to the dust components following the stellar distribution. Furthermore, in ETGs, it is heated solely by the old stellar populations, while in the discs of spiral galaxies, young stars contribute significantly to the heating of the diffuse dust \citep[see e.g. ][and references therein]{DeLooze2014,Bendo2015}. 

To explain the dust mass discrepancy found in ETGs, a diffuse dust component should be about 10-100 times as massive as the dust lane (based on \textit{IRAS} and \textit{ISO} mass estimates, respectively). Such a vast reservoir of cold dust should then be observable with the \textit{Herschel} Space Observatory \citep{Herschel}. However, in most ETGs, the FIR and sub-mm emission is much more compact \citep[see e.g. ][]{Smith2012}. As a result, dust emission is not spatially resolved in the vertical direction in Herschel observations. Therefore, a direct detection of a diffuse dust component following the stellar distribution has not yet been reported in ETGs.

We can now revisit these problems using high-quality data covering a broad range of wavelengths and advanced models. In the UV, optical and NIR, large surveys such as SDSS \citep{SDSS}, 2MASS \citep{2MASS}, \textit{GALEX} \citep{GALEX}, and \textit{WISE} \citep{WISE} have made multi-wavelength datasets available for many objects. In the FIR/sub-mm regime, \textit{Herschel} proved a tremendous leap forward in terms of sensitivity. Consequently, the dust content in ETGs has been studied intensely over the past few years. Dedicated investigations of individual targets (e.g. M87 \citealt{Baes2010}, M86 \citealt{Gomez2010}, and NGC 4125 \citealt{Wilson2013}) were supplemented with statistical studies. For example, from the \textit{Herschel} Reference Survey \citep[HRS, ][]{Boselli2010}: \citet{Smith2012,Cortese2012b}, from the \textit{Herschel} Virgo Cluster Survey \citep[HeViCS, ][]{Davies2010}: \citet{Clemens2010,diSeregoAlighieri2013}, and from the \textit{Herschel} Astrophysical Terahertz Large Area Survey \citep[H-ATLAS, ][]{Eales2010}: \citet{Agius2013,Rowlands2012}.
As the SPIRE \citep{SPIRE} and PACS \citep{PACS} instruments on-board the \textit{Herschel} spacecraft cover wavelengths from 70-500 $\mu$m, the peak of cold dust emission could be sufficiently sampled, and more reliable dust masses could be derived for all of these objects.

On the modelling side, significant progress has been made in the field of dust radiative transfer \citep[for a review, see][]{Steinacker2013}. Multiple efforts have already succeeded in fitting radiative transfer models to individual galaxies \citep{Xilouris1997,Xilouris1999,Popescu2000,Bianchi2007,DeLooze2012a,DeLooze2012b,DeGeyter2013,DeGeyter2014}. Up to now, this has mainly been applied to edge-on spiral galaxies, which inherently have morphologies that are more complex than ETGs. A similar way of modelling can thus be applied to ETGs with a prominent dust lane.

In this paper, we re-evaluate the dust mass discrepancy in ETGs using new high-quality optical and FIR/submm data, and advanced models. NGC 4370 is an edge-on lenticular in the Virgo cluster at a distance of 23 Mpc (taken from the GoldMine archive; \citealt{Goldmine}). It has been classified as a boxy S0 galaxy in several independent studies \citep{Binggeli1985,Baillard2011,Buta2015}. However, the edge-on view makes it difficult to classify because stellar discs or spiral arms may be hidden by the dust lane. A difference of more than a factor 3 between the dust mass obtained from the modelling of optical data and from IRAS fluxes was found for this galaxy \citep{Finkelman2008}. Since \textit{IRAS} dust masses are usually underestimations, this discrepancy can only widen. Thanks to its smooth stellar profile and conspicuous, regular dust lane, the galaxy is suited to several (semi-)analytical modelling techniques. Excellent optical data are now available for this object, complemented by deep FIR/sub-mm observations from \textit{Herschel}. We therefore choose NGC 4370 for this pilot study. The paper is organised in the following way. The data is described in Sect.~\ref{sec:data}, and our modelling efforts are outlined in Sect.~\ref{sec:results}. We discuss the results in Sect.~\ref{sec:discussion} and present our conclusions in Sect.~\ref{sec:conclusions}.

\section{Data} \label{sec:data}

\subsection{Optical/NIR data}

Optical imaging for NGC 4370 was obtained from the Next Generation Virgo Cluster Survey \citep[NGVS,][]{Ferrarese2012}. This legacy survey was executed with the MegaCam wide-field optical imager at the Canadian-France-Hawaii-telescope over the 2009A-2013A cycles. All fields were observed in the  $ugiz$ filters, with some also covered in the $r$ band. The survey consists of 117 separate MegaCam fields of $\sim$1.0 deg$^2$ with a pixel scale of $0.187$ arcsec. The fields overlap slightly, covering a total area of more than 104 deg$^2$ on the sky. The fields are  centred on M87 and M49, and they map the Virgo cluster out to one virial radius. The NGVS data is photometrically calibrated using stars from the SDSS resulting in rms errors of $0.01-0.02$ mag \citep{Gwyn2008}. Through dedicated data processing, a point-source depth of $\approx25.9$ mag ($10\sigma$) is achieved in the $g$ band. For extended sources, the surface brightness limit is $g = 29$ mag arcsec$^{-2}$ for a $2\sigma$ detection above the sky.

\begin{figure}
        \resizebox{\hsize}{!}{\includegraphics{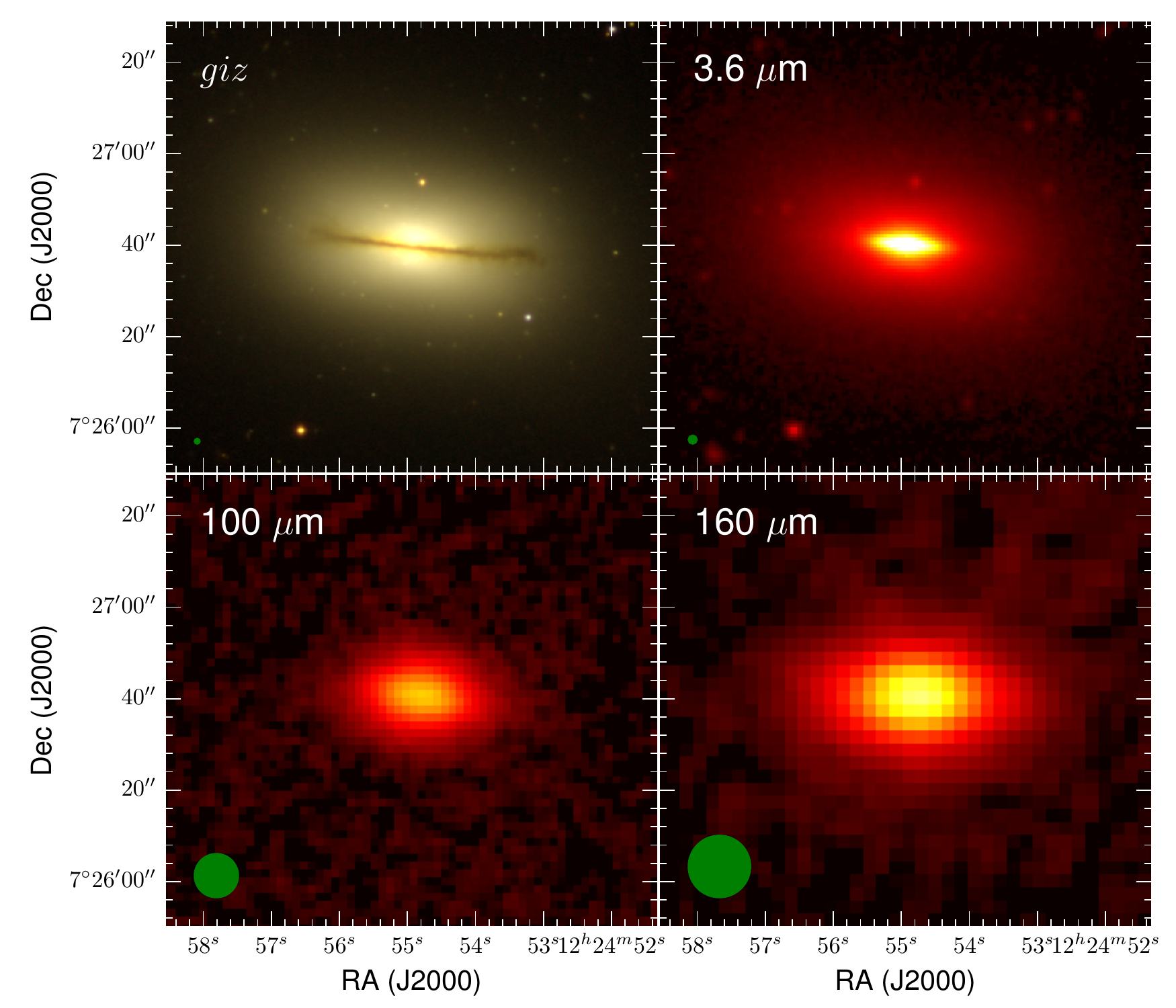}}
        \caption{Overview of the dataset. Top left: RGB colour map of NGC 4370 based on the $giz$ images of the NGVS. Top right: IRAC 3.6 $\mu$m image from the S$^4$G. Bottom: PACS 100 $\mu$m (left) and 160 $\mu$m (right) images. The beam size is indicated as a green circle in the lower left corner of each image.}
        \label{fig:data}
\end{figure}

NGC 4370 was imaged in the $ugi$ and $z$ bands. This dataset exposes the galaxy in great detail (seeing $< 1''$) and allows for an accurate study of the extinction effects (see Fig.~\ref{fig:data}). The fully reduced and calibrated images were first brought to the same point spread function (PSF). The full width half maximum (FWHM) of the PSF was determined for each band as an average of several brighter stars in the image. The resolution (FWHM) was measured to be 0.81, 0.87, 0.53, and 0.70 in the $u$, $g$, $i$, and $z$ bands, respectively, and therefore all images were convolved to a common 0.87 arcsec seeing using a Gaussian convolution kernel. 
The frames were regridded to the pixel grid of the $g$ band as well in order to work with a set of pixels that correspond to the exact same physical regions. After that, the images were rotated over $6 \degree$ in such a way that the dust lane fell along the x-axis of the frame. Both steps required some interpolations that reduced our working resolution to $0.98''$. This corresponds to 97 parsecs at the distance adopted for NGC 4370 (23 Mpc).

In a final step, bright foreground stars were masked out in each band. The stars were identified based on their shape and offset to the galaxy's brightness profile and verified with the NOMAD catalogue \citep{NOMAD}. Background galaxies in the field of view were identified using NED \footnote{http://ned.ipac.caltech.edu}. All reported galaxies in the field lie near the edge of the image, and the brightest one has a $g$ band magnitude of only 20.3. We therefore did not mask any background  sources.

The dust in NGC 4370 greatly affects the appearance of the optical images, which we use to our advantage. However, it is also useful to gain more insight into the underlying stellar morphology. We therefore include the IRAC 3.6 $\mu$m image from the Spitzer Survey of Stellar Structure of Galaxies \citep[S$^4$G, ][]{Sheth2010} which is shown in the top right-hand panel of Fig.~\ref{fig:data}. At these wavelengths, we expect minimal dust contamination and still benefit from a decent spatial resolution (FWHM $\sim$1.7 arcsec).

\subsection{FIR/sub-mm data}

The \textit{Herschel} Virgo Cluster Survey \citep[HeViCS;][]{Davies2010} observed 84 deg$^2$ of the Virgo Cluster with both PACS (100, 160 $\mu$m) and SPIRE (250, 350, 500 $\mu$m). The observations were done in parallel fast-scan mode (60 arcsec s$^{-1}$). The area is subdivided into four areas, corresponding to the region around M87 (V2), and the clouds to the north-west (V1), west (V3), and south (V4) of V2. Each area was covered by a $4\times4$ sq. degree tile and observed four times in two orthogonal scan directions to reduce the $1/f$ noise. The PACS data were reduced using a combination of HIPE \citep[v12.0,][]{Ott2010} and \texttt{Scanamorphos} \citep[version 24,][]{Roussel2013}. The latter, in particular, corrects for a flatfield distortion that causes an offset of up to $\sim$7$\%$ in the fluxes at 160 $\mu$m. In a subsequent observing run, 44 galaxies from the bright galaxy sample of HeViCS were observed with PACS in the 70 and 160 $\mu$m bands in map-making mode. NGC 4370 was one of them, and we use the 70 $\mu$m data from this run. These observations were reduced in a similar way to the parallel scan maps. The SPIRE observations were reduced using HIPE (v12.0) and the custom routine BriGAdE (Smith, in prep.) For more details on the data reduction, we refer the reader to \cite{Auld2013}. The final maps have a beam size (FWHM) of 5.7, 9.4, 13.4, 18.2, 25.4, and 36.0 arcsecs for 70, 100, 160, 250, 350, and 500 $\mu$m, respectively, corresponding to a physical scale of 0.64, 1.05, 1.49, 2.03, 2.83, and 2.01 kpc at a distance of 23 Mpc for NGC 4370.

Because NGC 4370 lies on the intersection between two tiles (V3 and V4), it was observed --partially-- by scans targeting the two different fields. It turns out that the galaxy was observed by a total of 13 scans, which were thus used to construct the maps at 100 and 160 $\mu$m. This is allowed by the option \texttt{cutout} in \texttt{Scanamorphos}, which forces the use of only those parts of the scanlegs that contain emission from the galaxy. This will enhance the signal-to-noise ratio (S/N), positively affecting the flux uncertainty determination, so we chose to redo the flux density measurement in the PACS 100 $\mu$m and 160 $\mu$m bands. We find new values of $F_{100\,\mu\mathrm{m}} = 3.03 \pm 0.18$ Jy and $F_{160\,\mu\mathrm{m}} = 3.65 \pm 0.24$ Jy. This is about $3 \%$ lower for $F_{100\,\mu\mathrm{m}}$ and $6\%$ for $F_{160\,\mu\mathrm{m}}$ compared to the measurements of \citet{Cortese2014}, who used the standard data products. These are, however, not corrected for the flatfield distortion in the PACS bands, while our measurements do include this correction. Figure \ref{fig:data} shows the newly reduced PACS maps. We also measure the integrated flux of the PACS 70 $\mu$m image and find $F_{70\,\mu\mathrm{m}} = 1.348 \pm 0.073$ Jy, which is fully consistent with the MIPS 70 $\mu$m flux measured by \citet{Bendo2012}. Since the SPIRE flux measurements from \citet{Ciesla2012} are already of good quality with just eight cross-scans, we take these for our further analysis. They are $F_{250\,\mu\mathrm{m}} = 1.95 \pm 0.05$, $F_{350\,\mu\mathrm{m}} = 0.77 \pm 0.04,$ and $F_{500\,\mu\mathrm{m}} = 0.25 \pm 0.02$ Jy. Table~\ref{tab:fluxes} summarises the integrated fluxes for NGC 4370 used in this work.

\section{Methods and results} \label{sec:results}

We explore several methods of investigating the extinction properties and determine the dust mass of NGC 4370. Each model treats the dust geometry, hence the corresponding attenuation effects, in a different way.
It is important to note that different dust mass estimates cannot be compared unless a common dust model is used. To achieve this consistency, we adopt the \cite{Draine2007} dust model throughout the paper. The dust model is derived from observations in our own Galaxy and consists of carbonaceous grains, amorphous silicates, and polycyclic aromatic hydrocarbons (PAH).

\subsection{Dust masses from FIR/sub-mm emission} \label{subsec:FIRmasses}

The bulk of the dust in galaxies resides in the diffuse ISM at low equilibrium temperatures of $10-30$ K. This component is directly measurable through its emission in the FIR/sub-mm. The emission is mostly optically thin and so guarantees that all the dust will be visible,  unlike in the case of the UV/optical/NIR where some dust may be hard to trace. Although the SPIRE observations were confusion limited, it may still be that we miss faint emission at larger radii. From the stacking of HRS spiral galaxies, Smith et al \textit{(in prep.)} find that $\sim$2.4$\%$ of the dust resides outside the D$_{25}$ radius. Assuming this is even less for ETGs, the missing dust will not contribute significantly to the sub-mm emission.

In the past, the dust mass and temperature of dusty ETGs were derived using a modified black body function \citep{Hildebrand1983} and \textit{IRAS} fluxes \citep[e.g.][]{Young1989,Goudfrooij1995,Patil2007,Finkelman2008}. For NGC 4370, in particular, a dust mass of $4.54_{-0.31}^{+0.33} 10^{5}\, \mathrm{M}_\odot$ was found, scaled to the distance we assume in this work (23 Mpc). The dust temperature from the \textit{IRAS} colours was found to be $T_\text{d} = 43.1 \pm 1.3$~K \citep{Finkelman2008}. It must be noted, however, that \textit{IRAS} was unable to detect dust with temperatures colder than $25$ K. The values thus only apply to the warm dust component.

\begin{figure}
        \resizebox{\hsize}{!}{\includegraphics{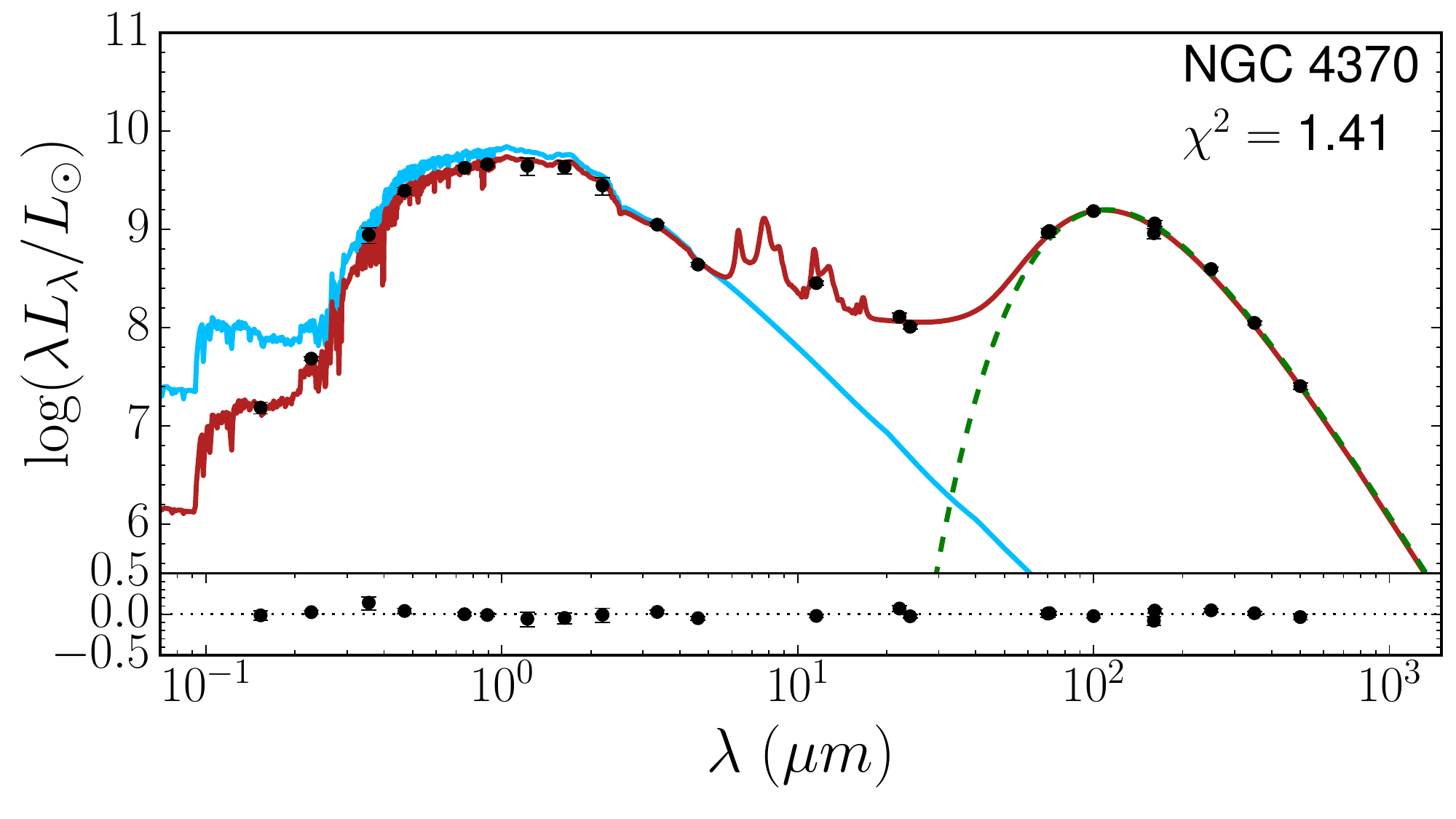}}
        \caption{MAGPHYS fit of the panchromatic SED of NGC 4370. The black points are the observed fluxes, the blue line is the unattenuated (stellar) SED, and the red line the attenuated, best-fitting template SED. The residuals are plotted below. The dashed green line is the best-fitting modified black body function to the \textit{Herschel} fluxes.}
        \label{fig:NGC4370_SED}
\end{figure}

Since the broad availability of \textit{Herschel} observations, it has become common practice to estimate the dust mass from a modified black body fit to the FIR and sub-mm fluxes. Following the recommendations of \citet{Bianchi2013}, we only fit a modified black body model with a fixed $\beta = 2$. This allows us to determine the dust mass using the \citet{Draine2007} dust model, with an emissivity of $\kappa_{350 \mu\mathrm{m}} = 0.192$ m$^2$\ kg$^{-1}$ (see Fig.~\ref{fig:NGC4370_SED}). We find a temperature $T_\text{d} = 21.8 \pm 0.3$~K and a total dust mass $M_\mathrm{d}^\text{modBB} = 6.17_{-0.36}^{+0.38} \times 10^{6}\, \mathrm{M}_\odot$. This is in line with \citet{diSeregoAlighieri2013}, who used similar data and methodology. It is roughly an order of magnitude higher than the \textit{IRAS}-derived dust mass and states again the importance of sampling the sub-mm emission of the cold dust component.

\begin{table}
\caption{Observed fluxes used for the SED fits for NGC 4370.}
\label{tab:fluxes}
\centering     
\begin{tabular}{>{\raggedright\arraybackslash}m{2.2cm}>{\centering\arraybackslash}m{2.5cm}>{\centering\arraybackslash}m{1.1cm}>{\centering\arraybackslash}m{1.1cm}}
\hline 
\hline
Band & Flux (mJy) & Source \\ [1ex]
\hline 
$FUV$                   & $0.0463 \pm 0.0063$   & a\\           
$NUV$                   & $0.2166 \pm 0.0059$   & a \\
$u$                             & $6.1 \pm 1.1$         & b \\
$g$                             & $22.9 \pm 1.8$        & b \\
$i$                             & $62.3 \pm 2.5$                & b \\
$z$                             & $80.9 \pm 3.2$                & b \\
$J$                             & $107 \pm 21$          & c \\
$H$                             & $138 \pm 21$          & c \\
$K$                             & $121 \pm 24$          & c \\
WISE $1$                        & $73.8 \pm 1.9$                & d \\
WISE $2$                        & $39.8 \pm 1.3$                & d \\
WISE $3$                        & $64.8 \pm 3.1$                & d \\
WISE $4$                        & $56.4 \pm 4.7$                & d \\
MIPS 24 $\mu$m  & $48.3 \pm     2.1$            & e \\
MIPS 70 $\mu$m  & $1284 \pm 132$                & e \\
PACS 70 $\mu$m  & $1348 \pm     73$             & f \\
PACS 100 $\mu$m & $3027 \pm 172$        & f \\
PACS 160 $\mu$m & $3647 \pm 231$                & f \\
MIPS 160 $\mu$m & $2880 \pm 350$                & e \\
SPIRE 250 $\mu$m        & $1947  \pm 53$                & f \\
SPIRE 350 $\mu$m        & $772   \pm 38$                & f \\
SPIRE 500 $\mu$m        & $251   \pm 19$                & f \\
\hline
\end{tabular}
\tablebib{
(a) \citet{Cortese2012a},
(b) NGVS, this work.
(c) GoldMine \citep{Goldmine},
(d) \citet{Agius2015},
(e) \citet{Bendo2012},
(f) HeViCS, this work.
} 
\end{table}

We also tested a more advanced spectral energy distribution (SED) fitting model: MAGPHYS \citep{daCunha2008}. This code allows simultaneous treatment of both stellar and dust components, preserving the energy balance between them. It has been used on a wide variety of galaxies \citep[e.g.][]{daCunha2010,DSmith2012,Clemens2013,Lanz2013,Viaene2014,Baes2014}. We sampled the panchromatic SED of NGC 4370 using \textit{GALEX} fluxes from \citet{Cortese2012a}, the $ugiz$ fluxes from the NGVS, $JHK$ magnitudes from the GoldMine archive \citep{Goldmine}, \textit{WISE} fluxes from \citet{Agius2015}, and MIPS fluxes from \citet{Bendo2012}. The \textit{Herschel} data described above complete our dataset. Table~\ref{tab:fluxes} lists the fluxes that were used. The best-fitting SED is shown in Fig.~\ref{fig:NGC4370_SED}, as is the intrinsic (stellar) SED without dust attenuation. 
The model is able to match the observed data remarkably well. If we use the same \citet{Draine2007} dust model, we find a total dust mass of $5.7^{+0.63}_{-0.07} \times 10^{6}\, \mathrm{M}_\odot$ and a temperature of $22.4_{-0.4}^{+0.1}$ K for the cold dust. It is reassuring that we retrieve --within the error bars-- a similar cold dust temperature and total dust mass to the modified black body model.

We note that even these dust mass determinations are still only a best guess of the true dust mass. There are known caveats in the modified black body fitting method. First, a single temperature is assumed for the cold, diffuse dust. In reality, this is likely a continuous range of temperatures, heated by a combination of emission sources. The multiple thermal components of dust appear to blend in such a way that it can be fit by a single modified black body \citep{Bendo2015}. In this scenario, the resulting dust temperatures from a single modified black body may be higher than the actual dust temperatures, and the resulting dust masses may be biased to lower values. Second, it is assumed that the dust in NGC 4370 has the same mixture as in our own Galaxy and that this mixture is the same everywhere in the galaxy. This should not,
however, affect relative dust mass differences such as those investigated in this paper. Last, recent results from \citet{PlanckXXIX2014} indicate that the \citet{Draine2007} dust model overestimates the dust mass by a factor 2-4. Throughout this work we nevertheless consider SED fitting to be the most objective and reliable way to estimate at least the order of the true dust mass in NGC 4370.

\subsection{Dust masses from colour maps} \label{subsec:colourmaps}

\begin{figure}
        \resizebox{\hsize}{!}{\includegraphics{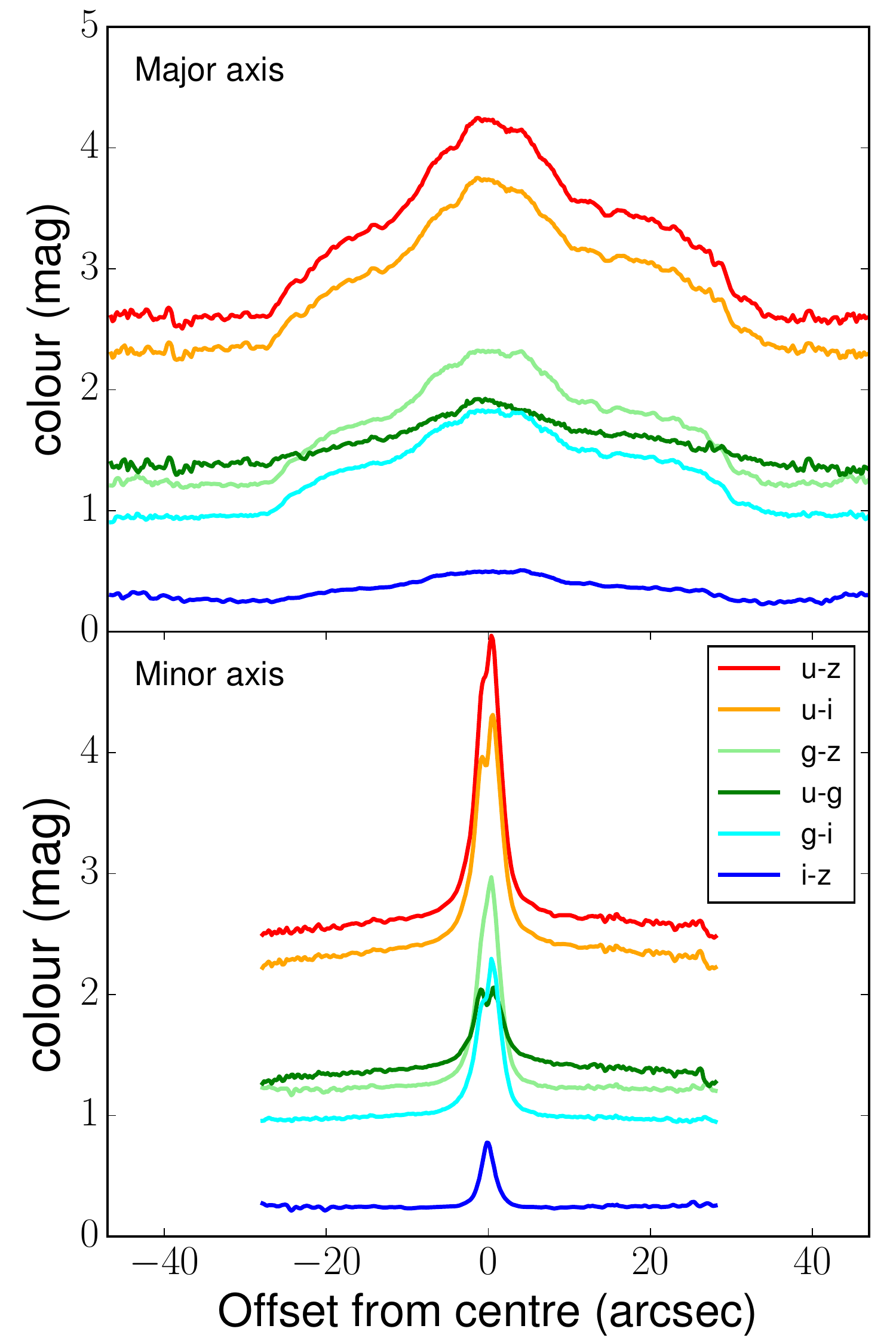}}
        \caption{Colour profiles along the major and minor axes, derived from each colour map of NGC 4370.}
        \label{fig:ColProfiles}
\end{figure}

Given the homogeneous dataset, colour maps are the simplest way to trace the extinction by dust. We can write the observed colour between two bands X and Y as 
\begin{equation}
(X - Y) = (X_0 - Y_0) + (A_X - A_Y)
\end{equation}
where $X_0 - Y_0$ is the intrinsic colour and $A_X - A_Y$  the difference in attenuation between the two wavelength bands, also called selective attenuation. The intrinsic colour $X_0 - Y_0$ is a manifestation of the wavelength-dependent luminosity of the stellar populations. The intrinsic colour was determined from the regions in the galaxy that do not show any obscuration by dust. Because we have four bands, $six$ colour maps can be constructed: $u-g$, $u-i$, $u-z$, $g-i$, $g-z$, and $i-z$. Figure~\ref{fig:ColProfiles} shows radial profiles along the major and minor axes for each of the colours. The same profile is retrieved in all colours: along the major axis, reddening starts increasing around $\sim$30 arcsec from the centre when dust reddening starts to be effective. It peaks in the centre where most dust is expected along the line of sight. Along the minor axis, a small gradient is present outside the dust lane. The reddening increases sharply when entering the dust lane at $\sim$4 arcsecs and peaks in the centre. Not surprisingly, the highest reddening is found for the $u-z$ colour and the lowest reddening for $i-z$.

\begin{figure}
        \resizebox{\hsize}{!}{\includegraphics{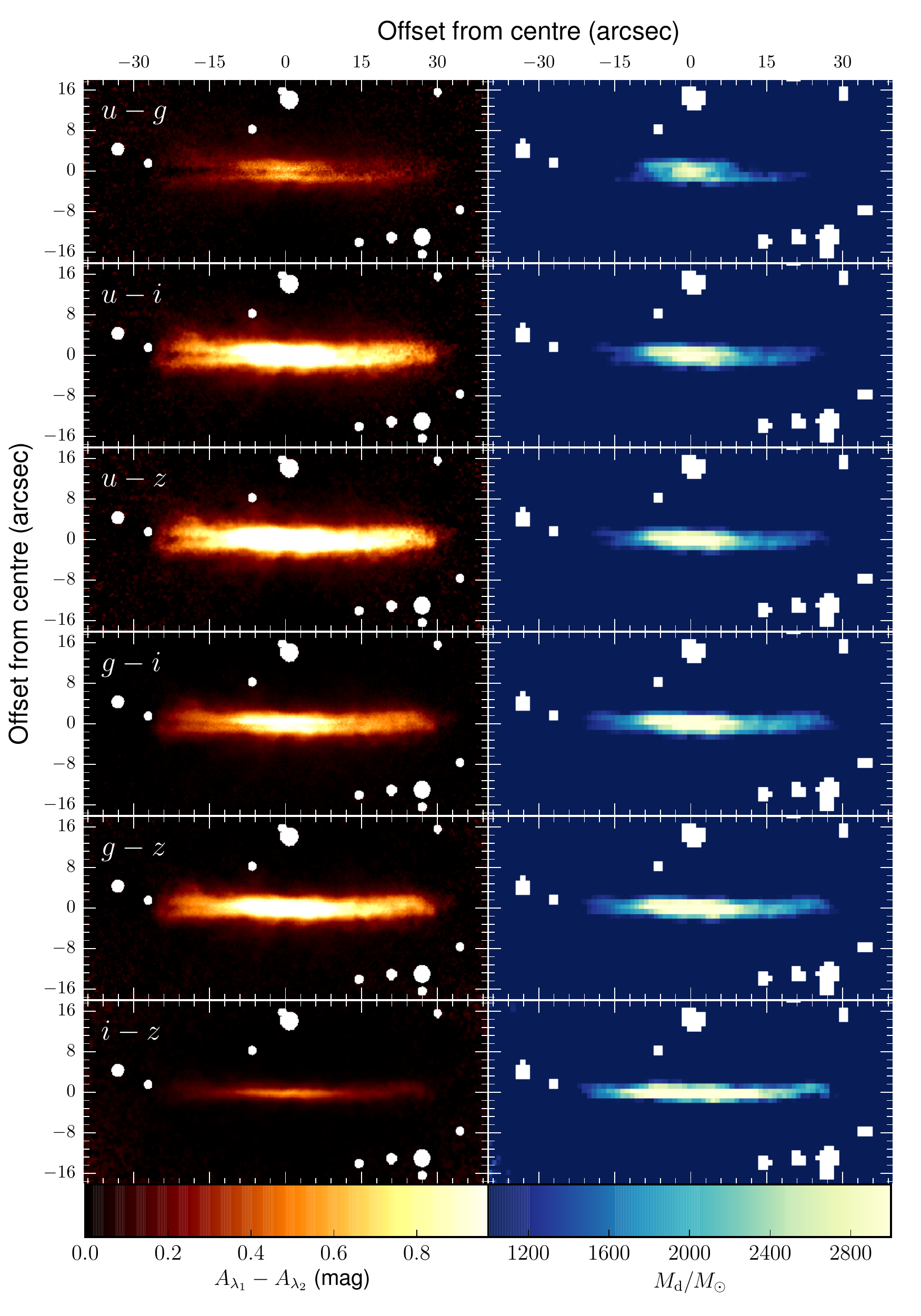}}
        \caption{Left: Selective attenuation ($A_{\lambda_1} - A_{\lambda_2}$) for all optical colours of NGC 4370. Right: corresponding dust mass maps in the foreground screen geometry at the critical 1 arcsec resolution. The white dots indicate masked foreground stars.}
        \label{fig:ColourMaps}
\end{figure}

Interestingly, at the centre of the dust lane, a minor dip is observed in the colour profiles along the minor axis. This effect is most evident in the $u-g$ profile (dark green), but also visible in most of the other profiles. Figure~\ref{fig:ColourMaps} shows the maps of selective attenuation, i.e. colour maps minus intrinsic colour. Here, the central part of the dust lane also shows this dip in reddening as a dark line running along the major axis. This \textit{\emph{colour inversion effect}} has been observed by \citet{Bianchi1999} in the B-I colour map of NGC 891, a massive edge-on spiral galaxy. It is suggested that saturation of the dust extinction occurs in these regions. At the same time, the number of stars between the dust and observer is expected to be largest here. As a result, the colour of the stars that lie in front of the dust becomes dominant over the reddening by dust. This makes the central part appear bluer than the dusty regions right above or below.

We must note that there are known colour gradients in ETGs, which possibly affect the measurement of the intrinsic colour. \citet{Petty2013} showed that the inner parts of ETGs are generally redder than the outskirts, and several spectroscopic studies have revealed age and metallicity gradients \citep[e.g.][]{Carollo1993,Kobayashi1999,Baes2007,Koleva2011}. As a result, when measuring the intrinsic colours of NGC 4370 outside the dust lane, it is possible that we pick up bluer colours than those behind the dust lane. This would result in slightly lower $A_X - A_Y$ values, hence lower dust masses. We confirm colour gradients along the minor axis of NGC 4370 (see Fig.~\ref{fig:ColProfiles}), however they are small with respect to the reddening in the dust lane. We therefore assume that the above effect will only cause a limited decrease in dust mass. The resulting maps of selective attenuation are shown in the left-hand column of Fig.~\ref{fig:ColourMaps}.

On the other hand, since NGC 4370 is classified as an S0 type galaxy, a stellar disc may be present, concealed by the dust lane. The IRAC 3.6 $\mu$m image (see Fig.~\ref{fig:data}) indeed shows a disc-like feature in the centre of the galaxy. If this obscured disc contains some ongoing star formation, it will be bluer than the intrinsic colours we measure outside the dust lane. Consequently, this can lead to underestimating the dust mass. It is nearly impossible to determine the intrinsic colour of a potential stellar disc from the edge-on perspective, which immediately illustrates the limitations of this simple approach. Fortunately, the stellar disc is relatively small ($<25''$ in diametre or $<0.25$D$_{25}$) and limited to the central regions of NGC 4370, so this effect may be relatively small. Nevertheless, this implies that resulting dust masses should be seen as lower limits to the true dust mass in NGC 4370.

The degree of the reddening is limited (1-4 magnitudes) because we only consider optical observations. The reason not to expand this set to UV or NIR observations is dual: first is the superb resolution in the optical images. The FWHM of the IRAC maps is twice that of the $g$ band, the largest optical beam size. For GALEX observations, the difference is more than a factor of 5. Second are the intrinsic colour gradients, which are much stronger when colour maps are created over these wide wavelength ranges. An accurate subtraction of the intrinsic colour is crucial for producing a reliable selective attenuation map. We therefore chose to proceed with the limited but fully consistent and high-quality optical dataset.

To convert (selective) attenuation maps to dust maps, we need to make assumptions about the star-dust geometry. Dust masses can be derived from the dust mass surface density in each pixel $\Sigma_d$. The total extinction coefficient $\kappa_\lambda$ of the dust at a certain wavelength, relates the dust mass to the optical depth $\tau_\lambda$:
\begin{equation} \label{eq:tautomass}
\tau_\lambda = \kappa_\lambda \Sigma_\mathrm{d}.
\end{equation}
To determine the optical depth from the attenuation maps, the geometry of the dust distribution needs to be taken into account. We consider two different dust geometries. The first one is a simple foreground dust screen between the stars and the observer:\\
\begin{equation} \label{eq:screen}
A_\lambda = -2.5\log(e^{-\tau_\lambda}) = 1.086\tau_\lambda.
\end{equation}
This geometry requires a minimal amount of dust to obtain the observed attenuation. The second model describes a layer of dust embedded in a layer of stars and is usually dubbed ``the sandwich model'' \citep{Disney1989,Boselli2003,Cortese2008}:
\begin{equation} \label{eq:sandwich}
A_\lambda = -2.5\log\left[ \left(\frac{1-\zeta_\lambda}{2}\right)\left(1+e^{-\tau_\lambda}\right) + \left(\frac{\zeta_\lambda}{\tau_\lambda}\right)\left(1-e^{-\tau_\lambda}\right)\right].
\end{equation}
The dimensions of the dust disc, relative to the stellar disc, are expressed by the parameter $\zeta_\lambda$. We limit ourselves to the two extreme cases of this geometry,  $\zeta_\lambda \rightarrow 1$ and $\zeta_\lambda \rightarrow 0$. The former limit corresponds to a dust layer of equal dimensions to the stellar layer. This boils down to a uniform mixing of dust and stars:
\begin{equation} \label{eq:mix}
A_\lambda = -2.5\log\left(\frac{1-e^{-\tau_\lambda}}{\tau_\lambda}\right).
\end{equation}
This formula can in principle be inverted analytically by means of the Lambert $W$ function to recover the optical depth from an observed attenuation \citep[see ][]{Boquien2013}.

On the other hand, when $\zeta_\lambda \rightarrow 0$, an infinitely thin slab of dust is embedded in the stellar layer along the line of sight:\begin{equation} \label{eq:embed}
A_\lambda = -2.5\log\left(\frac{1+e^{-\tau_\lambda}}{2}\right).
\end{equation}

Colour maps yield maps of selective attenuation $A_{\lambda_1} - A_{\lambda_2}$. This can be written as a combination of optical depth functions using equations \eqref{eq:screen} - \eqref{eq:embed}: $A_{\lambda_1} - A_{\lambda_2} = f(\tau_{\lambda_1}) - f(\tau_{\lambda_2})$. Using the \citet{Draine2007} dust model, the optical depth at one wavelength can be converted into the equivalent optical depth at another wavelength:
\begin{equation} \label{eq:taus}
\frac{\tau_{\lambda_1}}{\kappa_{\lambda_1}} = \frac{\tau_{\lambda_2}}{\kappa_{\lambda_2}}
\end{equation}
These equations were then numerically solved to yield a value for a specific optical depth, which in turn yields a dust mass through Eq.~\ref{eq:tautomass}.

Uncertainties on the attenuation and attenuation differences were derived following \citet{Goudfrooij1994}. Attenuation values were first derived on a pixel-by-pixel basis at the native pixel scale of $0.19''$. This is roughly five times smaller than the PSF of the convolved images ($0.98''$). We therefore rebinned each attenuation image using a $5\times5$ pixel box and  assigned to each (rebinned) pixel a value and uncertainty equal to the mean attenuation (or selective attenuation) and standard deviation calculated within each $5\times5$ pixel box. These uncertainties are propagated further through the standard bootstrapping method; we created 1000 randomly perturbed attenuation (or attenuation difference) maps based on the errors of the real attenuation maps. For each map, the resulting dust mass is then computed. The uncertainty on the  dust mass is then taken to be three times the standard deviation of the random set of dust masses. The final dust masses and their errors are listed in Table \ref{tab:dustmasses}.

Figure~\ref{fig:ColourMaps} shows the dust mass maps derived from the corresponding colour maps in the case of the foreground screen geometry. We only sum pixels inside the dust lane, because this is the only region where dust is directly visible. In the foreground dust screen geometry, the same morphology is retrieved for all dust mass maps.  Although there are some local variations in the dust mass per pixel, the total dust masses lie close to each other and average out to $5.60_{-1.3}^{+0.87} \times 10^5 M_\odot$, which should be interpreted as a lower limit. This is a value that is comparable to the \textit{IRAS}-derived dust mass, but still an order of magnitude below the dust mass estimates using \textit{Herschel} data. 

In the case of the sandwich model, retrieving a real solution for the optical depth was not possible for many pixels. For the $u-z$ colour, for example, the maximal attainable selective attenuation based on eq.~\eqref{eq:mix} is 1.2597. For eq.~\eqref{eq:embed}, this is only 0.3276. In contrast, the observed selective attenuation for the $u-z$ colour reached 2.49. Consequently, the pixels with the largest selective attenuation are incompatible both with an embedded thin slab of dust ($\zeta_\lambda \rightarrow 0$) and with a uniform mixing of dust and stars ($\zeta_\lambda \rightarrow 1$). Their mixed nature means that changing the properties of the dust distribution also changes the location of the stars relative to the dust. Because there will always be stars in front of and behind the dust, there are inherent limitations to the attenuation values along a line of sight. It appears that the sandwich model -- in the limits we explore here -- cannot achieve attenuation values that are high enough to explain the observed ones. We therefore chose not to pursue these geometries any further for the colour maps.

\subsection{Dust masses from multi-Gaussian expansion models} \label{subsec:MGEmod}

As a third method, we compare the observations with a dust free model of the galaxy in each waveband to construct monochromatic attenuation maps. A dust-free model is constructed through multi-Gaussian expansion (MGE) fitting. This technique was first developed by \citet{Emsellem1994} and consists of the summing of multiple 2-D Gaussian functions to mimic the brightness profile of elliptical galaxies. \citet{Cappellari2002} constructed a dedicated set of IDL routines to efficiently fit a set of Gaussians to an observed image.

It is crucial that the modelling is based on the dust-free part of the galaxy so that the dust lane is masked. We fixed the coordinates of the centre of the galaxy as determined from the $z$ band image (which is assumed to have the least dust contamination) using the MGE-IDL routine \texttt{find\_galaxy}. NGC 4370 has a boxy stellar distribution, not easily modelled by an elliptical profile. The MGE fitting software fortunately allows each Gaussian to have a different position angle, relative to the x-axis of the frame. A boxy brightness profile can be constructed by summing multiple 2-D Gaussian distributions with a slightly different angle.

MGE models were fitted to each of the four bands of our data set. A roughly equal effective area was used to constrain the free parameters. This corresponded to a S/N threshold of three in the $u$ and $g$ bands, and five in the $i$ and $z$ bands. The number of initial Gaussians is fixed for each MGE run. Allowing more Gaussians increases the number of free parameters, so that more complex brightness profiles can be constructed. On the other hand, if too many free parameters are available, peculiar and unphysical profiles may be found. To find a balance, we evaluated the non-reduced $\chi^2$ and shape of the brightness profile as a function of the number of free parameters until the best fit was obtained. 

\begin{figure*}
        \resizebox{\hsize}{!}{\includegraphics{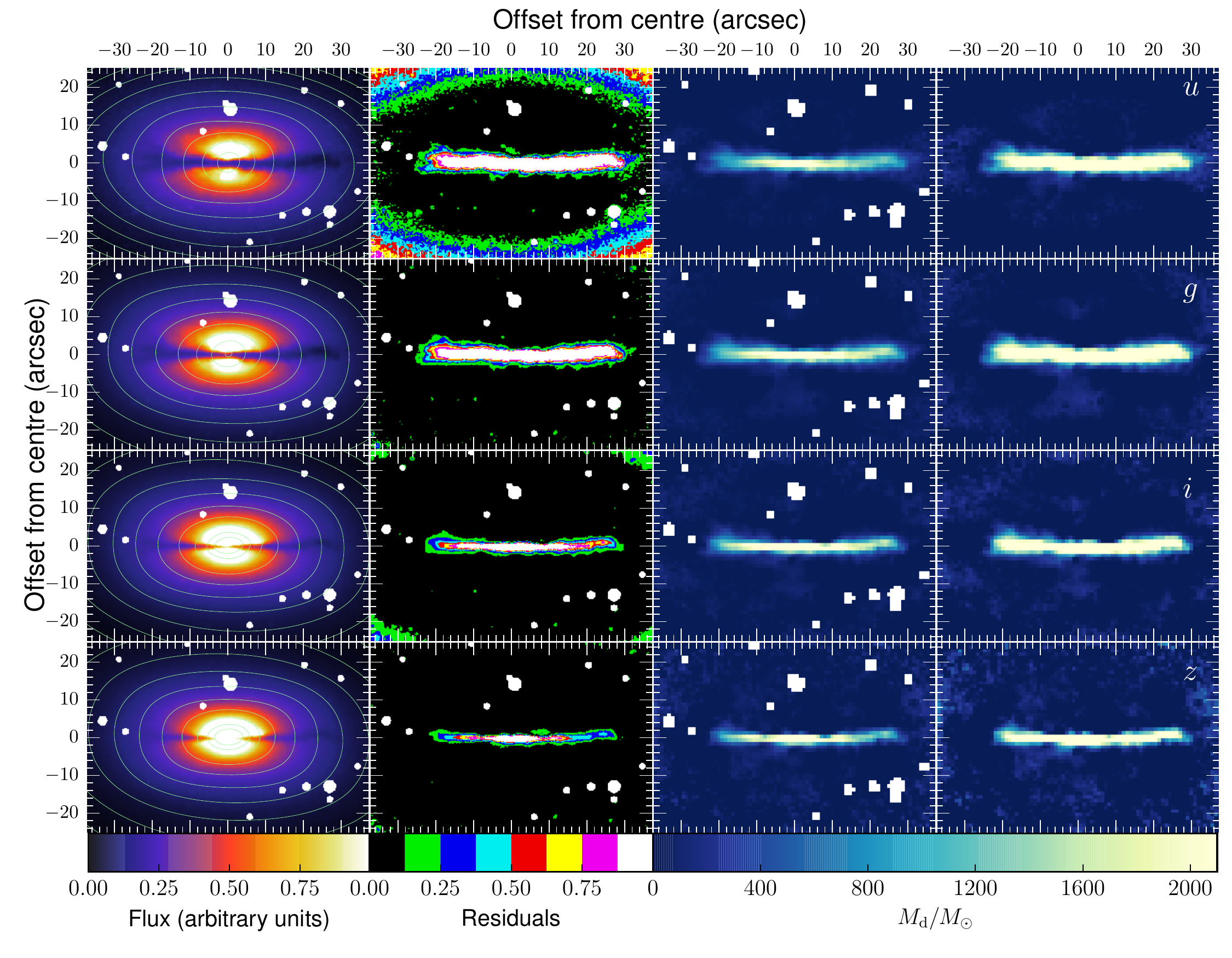}}
        \caption{First column: Observed images of NGC 4370, overlaid with contours of the corresponding dust-free MGE model. The contour levels are the same for all bands. 2nd column: Residuals of the MGE models and the observed images. 3rd column: corresponding dust mass maps in the foreground screen geometry and at the critical 1 arcsec resolution. 4th column: corresponding dust maps in the uniform star-dust mix geometry. The white dots indicate masked foreground stars.}
        \label{fig:MGEmodels}
\end{figure*}

The resulting set of Gaussians allows constructing a 2-D, dust-free model of the galaxy. We can now compute the attenuation $A_\lambda$ for each of the bands by taking
\begin{equation} \label{eq:mgeatt}
A_\lambda = -2.5\log\left(\frac{I_{\lambda\mathrm{,obs}}}{I_{\lambda\mathrm{,model}}}\right),
\end{equation}
where $I_{\lambda\mathrm{,model}}$ and $I_{\lambda\mathrm{,image}}$  are the model and reference frames, respectively. The attenuation values can then be converted to optical depths using equations \eqref{eq:screen}, \eqref{eq:mix}, and \eqref{eq:embed}. Equation \eqref{eq:tautomass} then allows the computation of the dust mass maps.

Figure~\ref{fig:MGEmodels} shows the results of the MGE modelling for the $ugiz$ bands. In the first column, the observations are overlaid on the contours of the dust-free MGE model. In all bands, the contours closely follow the stellar emission in the observed images. Generally, the model is elliptical in the centre and becomes boxier in the outer part of the galaxy. Owing to the obscuration of the dust lane, some small differences in the central contour shapes are visible across the wavebands. The second column shows the residuals of the MGE model and the observations. Obviously, the dust lane sticks out here because no dust was included in the MGE model. Apart from that, the residual levels are usually below $12.5 \%$ and only increase in the outer, low S/N regions. Using this formalism, it is thus possible to reproduce the stellar emission of NGC 4370, and we can now construct attenuation maps. These are shown in Fig.~\ref{fig:FitSKIRTatts} and discussed in Sect.~\ref{subsec:attmaps}.

In the last two columns of Fig.~\ref{fig:MGEmodels}, the resulting dust mass maps are displayed for the screen, along with uniform-mix geometries. In the case of the embedded thin slab (eq.~\eqref{eq:embed}), again no real solutions were found for many of the pixels. This particular geometry can only generate moderate attenuation values up to $2.5\log 2 = 0.753$, which are lower than most pixels in the dust lane. We therefore do not pursue this geometry any further.
For the other two geometries, the dust mass maps have similar morphology and pixel values. A slight decrease in pixel dust-mass values is observed in the $z$ band. This is not entirely surprising since the $z$ band is the longest in wavelength, hence least sensitive to attenuation by dust. Comparing both the dust screen (3rd column) and uniform dust mix (4th column) geometries, the latter clearly has higher dust masses in each pixel. Owing to the mixing of dust and stars, more dust is needed to obtain the same level of attenuation as for the foreground screen of dust. The obtained dust masses are consistent with this view. The total masses are listed in Table \ref{tab:dustmasses} and were obtained by summing the pixel values in the dust lane. As already visible from the maps, the dust masses in the $u$, $g$, and $i$ bands lie close to each other, within each geometry. The $z$ band total dust masses are significantly lower.

On average, the foreground screen yields a total dust mass of $3.35_{-0.90}^{+0.64} \times 10^5 M_\odot$ and the uniform mix geometry $7.9_{-2.5}^{+1.9} \times 10^5 M_\odot$. These values are comparable to the average total dust mass from the colour maps and consistent with the \textit{IRAS}-based dust mass for this galaxy. In the end, with this high-quality data, we are already able to alleviate the previously found dust mass discrepancy between optical and \textit{IRAS}-based dust masses. Of course, \textit{IRAS} masses are still insensitive to cold dust. Indeed, as for the colour maps, the masses are still about an order of magnitude lower than the ones obtained using the \textit{Herschel} observations, which do observe emission from cold dust.

\begin{table}
\caption{Dust masses for NGC 4370 as derived from the various methods, in units of $10^5 M_\odot$. The embedded slab geometry did not yield real solutions for either the colour or the MGE method and are therefore not included in the table.}
\label{tab:dustmasses}
\centering     
\begin{tabular}{>{\raggedright\arraybackslash}m{2.5cm}>{\centering\arraybackslash}m{2.0cm}>{\centering\arraybackslash}m{2.0cm}}
\hline
\hline
Colour maps             & Screen & Mix \\
$u-g$                   & $4.48 \pm 0.15  $ & no solution \\
$u-i$                   & $5.37 \pm 0.15 $ & no solution \\
$u-z$                   & $5.48 \pm 0.14 $ & no solution \\
$g-i$                   & $6.03 \pm 0.17  $ & no solution \\
$g-z$                   & $6.06 \pm 0.17 $ & no solution \\
$i-z$                   & $6.20 \pm 0.27 $ & no solution \\
Average                 & $5.60_{-1.3}^{+0.87}$ & no solution \\
\hline
MGE fits                        & Screen & Mix\\
$u$                             & $3.25 \pm 0.12$ & $8.20 \pm 0.31$ \\
$g$                             & $3.83 \pm 0.15$ & $9.36 \pm 0.37$ \\
$i$                             & $3.58 \pm 0.24$ & $8.14 \pm 0.55$ \\
$z$                             & $2.72 \pm 0.27$ & $5.93 \pm 0.49$ \\
Average                 & $3.35_{-0.90}^{+0.64}$ & $7.9_{-2.5}^{+1.9}$ \\
\hline
 & \multicolumn{2}{c}{FIR/sub-mm}\\
\textit{IRAS}   $^\mathrm{a}$   & \multicolumn{2}{c}{$4.54_{-0.31}^{+0.33}$} \\
ModBB           & \multicolumn{2}{c}{$61.7_{-3.6}^{+3.8}$} \\
MAGPHYS                         & \multicolumn{2}{c}{$57_{-1}^{+6}$} \\
\hline
 & \multicolumn{2}{c}{FitSKIRT} \\
Exponential disc        & \multicolumn{2}{c}{$50 \pm 24$} \\
Ring                            & \multicolumn{2}{c}{$33 \pm 27$} \\
IRAC Ring               & \multicolumn{2}{c}{$70 \pm 18$} \\
\hline
\end{tabular}
\tablebib{
(a) \citet{Finkelman2008}
} 
\end{table}

\subsection{Dust masses from radiative transfer modelling}

The most advanced way of treating the extinction of starlight through a dusty medium is through radiative transfer (RT) simulations. This technique allows more realistic dust distributions (e.g. embedded discs or rings) and takes the 3-D geometry of the problem
into account. The main advantage over the previous techniques is the self-consistent treatment of dust extinction along the line of sight. This provides a more realistic determination of the amount of dust needed to produce the observed dust lane.
Ideally, we would like to solve the inverse RT problem, i.e. find the best fitting parameters for a RT-model to match the observations. FitSKIRT \citep{DeGeyter2013, DeGeyter2014}, an optimization code built around the RT code SKIRT \citep{Baes2003,Baes2011,Camps2015}, is able to probe this parameter space efficiently and find the best-fitting model to a set of observations. FitSKIRT's optimization technique is based on genetic algorithms, which start with an initial population of 150 different parameter sets. For each set, a RT simulation is run and the simulated images are compared to the observed images. Then,  a second generation of 150 parameter sets is constructed from the best-fitting sets. Again, a RT simulation is run for each parameter set. We chose to limit the number of generations to 100, although convergence is usually reached much faster. The code was originally designed and tested to fit the complex geometries of edge-on spiral galaxies. These objects are successfully modelled with up to $19$ free parameters. Several of those describe the complex bulge-disc geometry of the stars. In our model, however, we keep the stellar geometry fixed, which drastically reduces the number of free parameters to fit, thereby increasing our confidence in the results of the fits.

Imaging in blue bands is ideal for investigating the extinction properties of dust, but it hides information on the underlying stellar distribution. The red bands, on the other hand, provide more accurate information on the stellar distribution, but it makes it more difficult to study the dust-extinction effects. A combined fit to all bands can alleviate these degeneracies, as shown by \citet{DeGeyter2014}. This procedure is called oligochromatic RT fitting. An additional benefit is the more stable convergence towards the optimal solution, particularly for low S/N observations \citep{Haussler2013,Vika2013}. We explore RT models with two different geometries for the dust distribution in NGC 4370: an exponential disc geometry and a ring geometry. Their best-fitting parameters are determined using FitSKIRT. We simultaneously fit each RT model to the four images ($u$, $g$, $i$, and $z$ bands).

The stellar distribution for each model is the same and derived from the $z$-band image. The longest waveband was chosen to limit contamination by the dust. We construct a dust-free MGE model similar to the one obtained in section \ref{subsec:MGEmod}. In this case, however, we force each Gaussian to have the same position angle $\theta = 90\degree$. Allowing different position angles for each Gaussian would severely increase the difficulty of deprojecting the 2-D model to a 3-D stellar distribution. We have chosen to work with only one position angle. The inclination angle for deprojection is left as a free parameter within the $88\degree < i < 92 \degree$ interval.

Both geometries are fitted five times, each time with a different random seed for the genetic algorithms, hence leading to a different solution. The variation in the resulting parameters of each individual fit is used as an indication of the uncertainty and the ability of FitSKIRT to constrain them. The difference between the models lies in the adopted dust distribution, which we discuss below. The results of the fitting are summarised in Table \ref{tab:FitSKIRT}, which lists the parameters of the dust geometries, and Table \ref{tab:dustmasses}, which lists the corresponding dust masses. 

\subsubsection*{Model A}

The first geometry consists of a classic exponential dust disc, generally applied in the RT models of edge-on spiral galaxies \citep{Xilouris1999,Popescu2000,Bianchi2007,Baes2010,Popescu2011,DeLooze2012b,DeGeyter2013,DeGeyter2014} and also derived from several FIR observations \citep[see e.g.][]{MunozMateos2009, Verstappen2013}. The geometry is determined by three free parameters: the radial scale length $h_R$, the axial scale height $h_z$, and the central dust density $\rho_0$. The dust density $\rho_d$, is then described by
\begin{equation}
\rho_d(R,z) = \rho_0\, \exp\left[-\frac{R}{h_R}-\frac{|z|}{h_z}\right],
\end{equation}
from which the total dust mass can be derived.

\begin{figure*}
        \resizebox{\hsize}{!}{\includegraphics{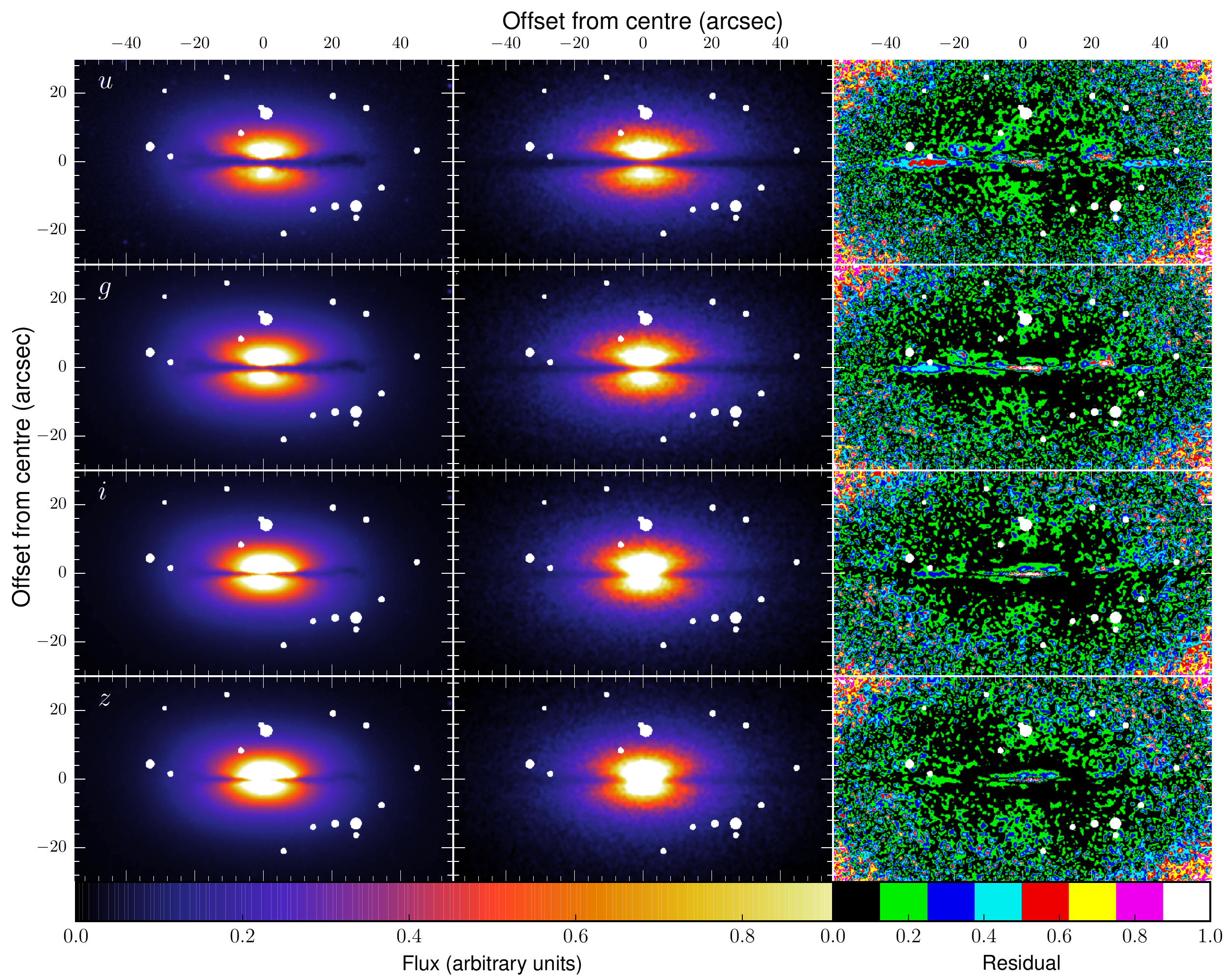}}
        \caption{FitSKIRT model results for the exponential dust disc geometry. Left column are the observations, with masked-out foreground stars. Middle column shows the corresponding model images. The residuals are shown in the right column.}
        \label{fig:DiscResults}
\end{figure*}

The observed images are reproduced relatively well (see Fig.~\ref{fig:DiscResults}), with residuals (right column) generally staying below $30 \%$. The noise in the residual images is a combination of observational noise and the Monte Carlo noise from the RT simulations. This is also why the models (middle column) appear noisier than the observations (left column). The dust lane in the model does seem to extend up to larger distances, while it falls off more abruptly in the observed images. In the central part, there is a difference of up to a factor of 2 in luminosity. The model slightly overestimates the luminosity in the nucleus of the galaxy in favour of a better fit to the outer regions. Since the S/N in the centre is highest, these pixels will have the most weight in the fitting algorithm. A sufficient number of well-fitting pixels with lower S/N are needed to compensate for a deviation in the model from the observations in these high-S/N pixels to still yield a lower $\chi^2$. NGC 4370 is not perfectly symmetric, as is already evident from Fig.~\ref{fig:data}. 

The western side of the galaxy shows a thicker dust lane with stronger extinction than the eastern side. On the other hand, the exponential dust disc is an axially symmetric analytical geometry. Therefore, it is not surprising that some asymmetry can be spotted in the residuals. It appears that the FitSKIRT optimal solution favours fitting the western part of the dust lane. As a result, the model overestimates the extinction in the eastern part by about $50 \%$ in the $u$ band and $20 \%$ in the $z$ band. The spread on the five independent fit results is not more that $10 \%$ except for the dust scale height and the dust mass, where it amounts to $50 \%$. We find a total dust mass of $(5.0 \pm 2.4) \times 10^6 M_\odot$ for this model.
A truncated exponential disc could potentially produce better fits. However, it would not change the total dust mass significantly because the mass in the outer parts is low due to the exponential fall off. Furthermore, the central over-density will persist. We therefore chose not to pursue a fit with these additional free parameters, but explored the possibility of a ring geometry.

\subsubsection*{Model B}

Dust could be distributed in a ring-like geometry. This has been observed for several ETGs \citep{Goudfrooij1994b,Bendo2006,Patil2007,Finkelman2008}, but is difficult to confirm through observations alone when seen edge-on. We explored a modified ring distribution, which has a Gaussian profile in the radial direction and an exponential fall-off in the vertical direction. It is characterised by four free parameters: the radius $R_0$ at which the highest density occurs, the radial width $W$ of density peak, the vertical scale height $H$, and again the dust density in the centre $\rho_0$. The density in the dust ring is described by
\begin{equation}
\rho_d(R,z) = \rho_0\,\exp\left[ -\frac{(R-R_0)^2}{2W^2} \right] \exp\left( -\frac{|z|}{H} \right)
.\end{equation}

\begin{figure*}
        \resizebox{\hsize}{!}{\includegraphics{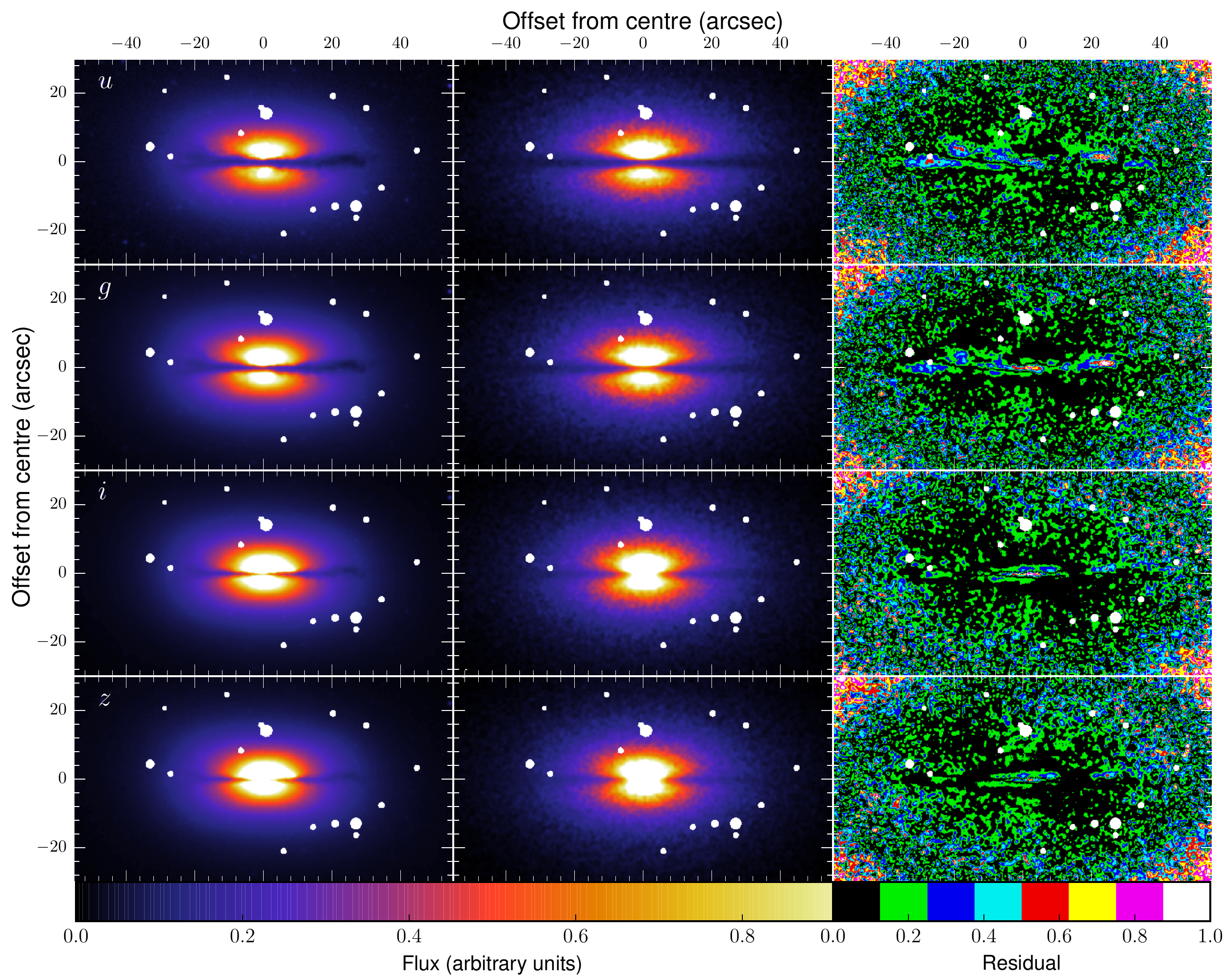}}
        \caption{FitSKIRT model results for the ring dust geometry. Left column are the observations with masked-out foreground stars. Middle column are the corresponding model images. The residuals are shown in the right column.}
        \label{fig:RingResults}
\end{figure*}

The observed images are also accurately reproduced by the dust ring geometry (see Fig.~\ref{fig:RingResults}). The spread on the five independent fit results is slightly greater than for the disc geometry. Especially the width $W$ and dust mass have large variations (up to a factor of 2.5). Nevertheless, the five independent fits yield residual maps that show very similar morphology. There is still a slight enhancement visible in the centre of the residuals, which is, however, less prominent than in Model A, and the discrepancy is not more than $50 \%$ in most pixels. This is also an axisymmetric geometry, and again the extinction in the eastern part of the galaxy is overestimated, while the western part is reproduced better. Judging from the low residual values in the outer parts of the dust lane ($30 - 50 \%$), Model B is more accurate than Model A in these regions. This is also visible in the model images, where the dust lane is more compact than Model A and corresponds better to the observations. Based on these considerations, a dust ring seems more likely than an exponential dust disc in NGC 4370, although the results lie close to each other. The total dust mass for this model is $(3.3 \pm 2.7) \times 10^6 M_\odot$ which is lower than, but consistent with Model A.

\subsubsection*{Attenuation maps} \label{subsec:attmaps}

\begin{figure*}
        \resizebox{\hsize}{!}{\includegraphics{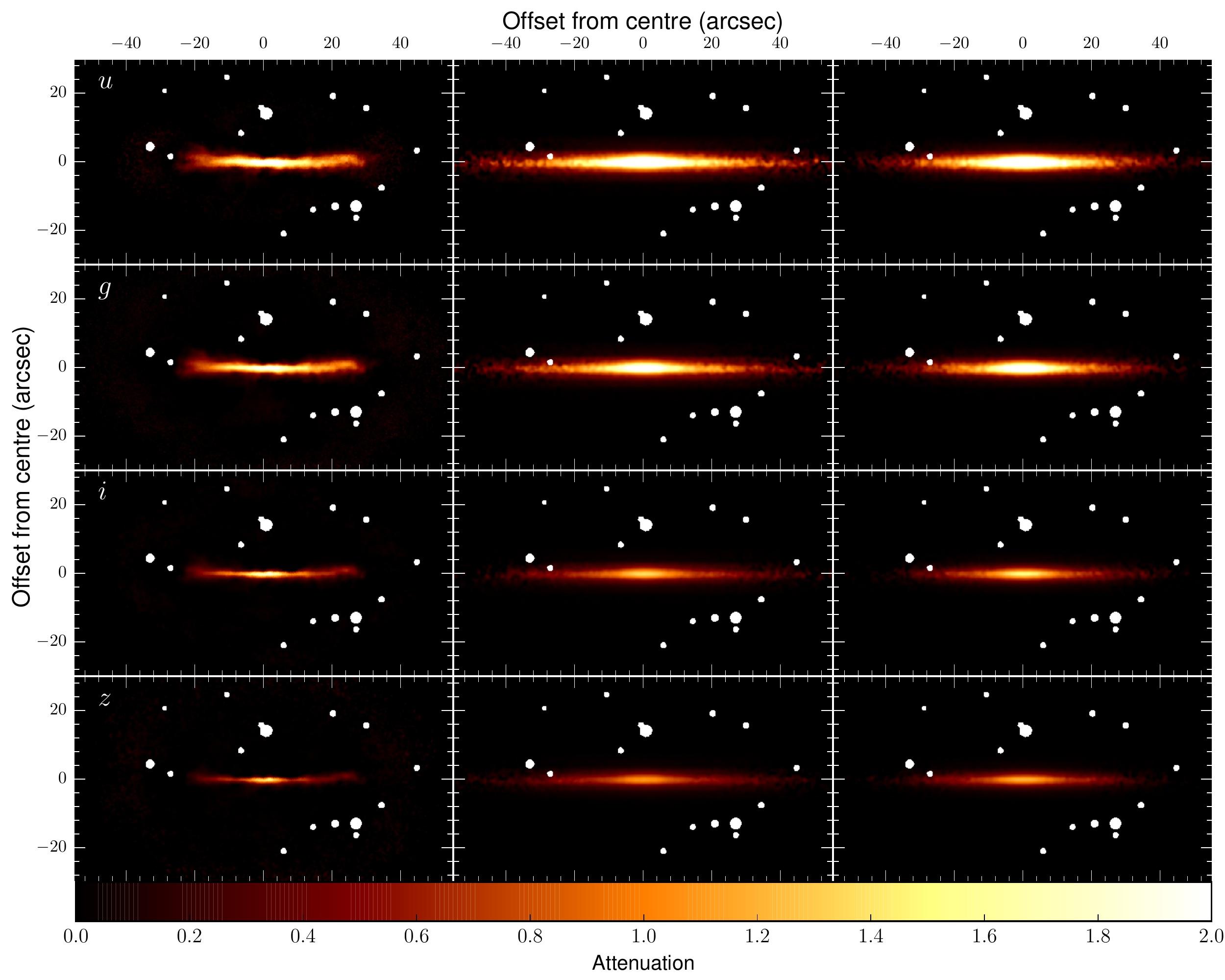}}
        \caption{Attenuation maps for the various models of NGC 4370. Left: Attenuation maps from the MGE models. Middle: maps from the FitSKIRT exponential dust disc model. Right: maps from the FitSKIRT dust ring model. All images are convolved to the same beam size.}
        \label{fig:FitSKIRTatts}
\end{figure*}

To investigate the validity of the FitSKIRT modelling further, we produced the attenuation maps for the $ugiz$ bands. SKIRT offers the option of also outputting a transparent model (i.e. only stellar emission), which serves a dust-free model. These were then convolved to the same PSF as the observed images and compared to the FitSKIRT best-fitting images using eq.~\eqref{eq:mgeatt}. Figure~\ref{fig:FitSKIRTatts} shows the resulting attenuation maps, together with the ones obtained from the MGE fitting in section \ref{subsec:MGEmod}. Attenuation maps offset the dusty and dust-free images and consequently enhance deviations from a smooth dust distribution. In the case of the FitSKIRT attenuation maps, which have smooth distributions as an input, this is limited to the Monte Carlo noise inherent to the RT simulations. For the MGE modelling, which works with the observed images as input, observational noise is added on top of the intrinsic deviations of a smooth dust distribution. With this in mind, we limit ourselves to a qualitative comparison of the attenuation maps. 

The MGE attenuation maps appear thin, but with constant thickness. At the edges, the attenuation seems to be slightly puffed up in the vertical direction. There is also a clear decrease in attenuation going from $u$ to $z$, which is inherent to the extinction properties of dust. The FitSKIRT attenuation maps exhibit the same decrease with increasing wavelength. There are some evident differences between the MGE maps and the FitSKIRT maps. In the case of the exponential disc geometry, the attenuation features are more elongated and are thicker in the middle than at the edges (as is inherent to the analytical distribution). As a result, they overestimate attenuation values in the centre and outskirts of the galaxy. The attenuation maps from the ring geometry lie closer to the ones from the MGE models. They are more compact in scale length and less peaked in the centre. Still, the ring geometry yields higher attenuation values in the centre and fails to reproduce the puffed-up morphology in the outer parts. 

In general, the radiative transfer models yield an attenuation map that has higher values (by an average factor of 2), but that resembles the observed map. This can partially contribute to the discrepancy in dust mass with the MGE-based estimates. However, all models produce roughly the same thin and elongated morphology. SKIRT treats the extinction of dust in a self-consistent way, so it is reassuring that the attenuation maps are reproduced this well. This boosts confidence in the applied method and the resulting dust mass parameters. It also underlines the importance of the dust geometry and a proper treatment of absorption and multiple scattering to convert these attenuation values to more reliable dust masses.

\subsubsection*{Effect of a stellar disc} \label{subsec:IRACfits}

As already mentioned in section \ref{subsec:colourmaps}, NGC 4370 may have a nuclear stellar disc. This is obscured by the dust lane in the optical images, but visible in the IRAC 3.6 $\mu$m image (see Fig.~\ref{fig:data}). Such a component is not explicitly included in the RT models when deriving the stellar distribution from the $z$ band. We explored the possibility of deriving the stellar distribution from the IRAC 3.6 $\mu$m image using an MGE model, similar to the one derived from the $z$ band image. The drawback to this method is the difference in spatial resolution between IRAC and the NGVS data. We only explored the ring geometry and ran the FitSKIRT simulation five times to get a grip on the uncertainties.

\begin{figure*}
        \resizebox{\hsize}{!}{\includegraphics{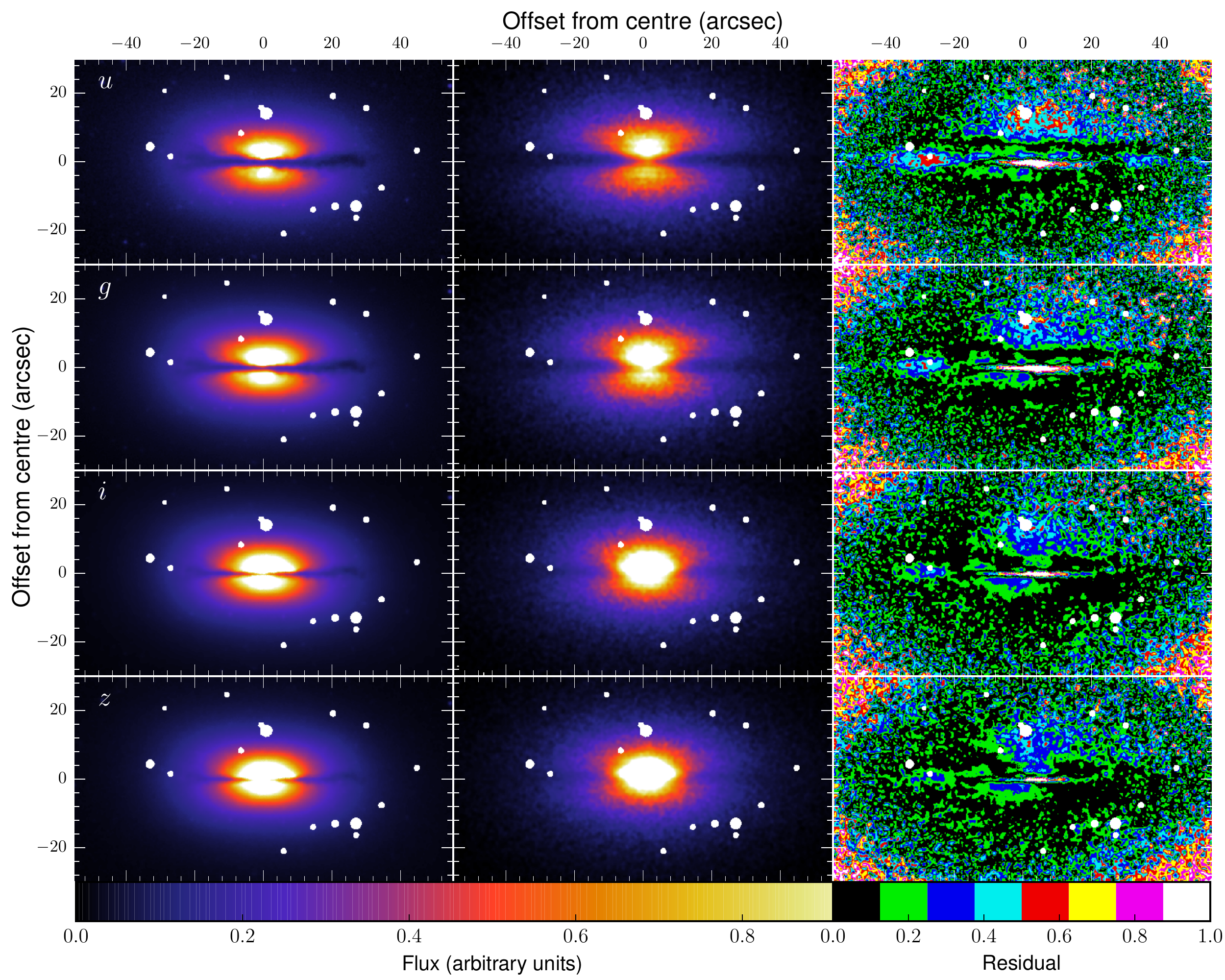}}
        \caption{FitSKIRT model results for the ring dust geometry, but with the stellar geometry derived from the IRAC 3.6 $\mu$m image. Left column shows the observations with masked-out foreground stars. The middle column gives the corresponding model images. The residuals are shown in the right column.}
        \label{fig:IRACRingResults}
\end{figure*}

Figure~\ref{fig:IRACRingResults} shows the results of the FitSKIRT simulation. At first sight, the observations are reproduced quite well. The dust lane is rather compact, and the outer stellar halo is retrieved. The inner stellar distribution, however, is much brighter above the dust lane in the model. Below the dust lane, it is fainter. The larger beam size of the IRAC image  also leaves its mark: the stellar component is generally more stretched along the major and minor axes.
The quality of the fit is easier to judge from the residuals, as shown in the right-hand column of Fig.~\ref{fig:IRACRingResults}. There is a large residual feature above the dust lane due to the overestimation of the stellar emission. The residuals at the location of the dust lane itself again show the features that are present for the other models (offsets in the centre and asymmetry in the outer part), but they are more prominent here. Just above and below the dust lane, the residuals of the nuclear stellar disc are also visible. This means the brightness profile of the stellar disc in the $ugiz$ bands does not scale linearly with the shape it has at 3.6 $\mu$m. It is less prominent in the optical bands. This suggests a colour gradient in the disc, which is stronger than the general stellar distribution. Consequently, if the disc is relatively red, the embedded star formation will be low. This would mean that the caveat we described at the end of section \ref{subsec:colourmaps} will have a relatively low impact on the dust mass estimate from the colour maps or the MGE fits. Thanks to the inclusion of a stellar disc, we retrieved a higher total dust mass of $(7.0 \pm 1.8) \times 10^6 M_\odot$, which still resembles the other FitSKIRT models. The dust masses of all models are consistent with each other within 1 rms. This again suggests that the effects of the nuclear stellar disc on the estimate of the dust mass are relatively small.

\begin{table}
\caption{Overview of the resulting dust geometry parameters and their variation from the FitSKIRT modelling of NGC 4370. The obtained dust masses are listed in Table \ref{tab:dustmasses}.}
\label{tab:FitSKIRT}
\centering     
\begin{tabular}{>{\centering\arraybackslash}m{3cm}>{\centering\arraybackslash}m{1cm}>{\centering\arraybackslash}m{3.0cm}}
\hline
\hline
Parameter & Unit & \textit{ugiz} $\pm$ rms \\
\hline
\multicolumn{3}{l}{Exponential disc} \\
\hline
Scale length     $h_R$& pc & $ 3966 \pm 1953$ \\
Scale height     $h_z$& pc & $ 155 \pm 17$ \\ 
Inclination     $i_{\mathrm{disc}}$& deg        & $ 89.27 \pm 0.10 $ \\
\hline
\multicolumn{3}{l}{Ring ($z$ band)} \\
\hline
Radius $R_0$            & pc    & $ 1969 \pm 498 $ \\
Width $W$               & pc    & $ 2130 \pm 3346 $ \\
Height $H$              & pc    & $ 146 \pm 14$ \\
Inclination $i_{\mathrm{Ring}}$ & deg           & $ 88.83 \pm 0.59 $ \\
\hline
\multicolumn{3}{l}{Ring (IRAC 3.6 $\mu$m)} \\
\hline
Radius $R_0$            & pc    & $ 1094 \pm 2085 $ \\
Width $W$               & pc    & $ 4199 \pm 1195 $ \\
Height $H$              & pc    & $ 299  \pm 93$ \\
Inclination $i_{\mathrm{Ring}}$ & deg& $ 89.08 \pm 0.04 $ \\
\hline
\end{tabular} 
\end{table}

\section{Discussion} \label{sec:discussion}

The existence of a diffuse dust component in dust-lane ETGs was first proposed by \citet{Goudfrooij1995} to explain the dust mass discrepancy between optically derived dust masses and \textit{IRAS}-based masses. The foreground dust screen geometry they adopted naturally yields lower limits to the optical dust mass. In a similar two-component modelling, \citet{Kaviraj2012} computed the dust mass in clumpy or filamentary structures in dusty ETGs. The dust mass in clumps was found to be on average five times lower than the \textit{IRAS}-derived total dust masses, again favouring a diffuse dust component, invisible in the UV/optical/NIR. A similar discrepancy in optical and IRAS-derived dust mass was found for NGC 4370 \citep{Finkelman2008}. In this work, we re-evaluate this discrepancy using data of higher quality at optical and FIR/sub-mm wavelengths. Using colour maps, we find an average dust mass of $5.6_{-1.3}^{+0.9} \times 10^5 M_\odot$ for a foreground screen geometry, which should be seen as a lower limit. From MGE modelling, we find $M_d = 3.4_{-0.9}^{+0.7} \times 10^5 M_\odot$ for the same geometry, and $M_d = 7.9_{-2.5}^{+1.9} \times 10^5 M_\odot$ for a uniform dust-star mix. On the other hand, modelling of the FIR/sub-mm SED yields $M_d = (6.2 \pm 0.4) \times 10^6 M_\odot$, and a panchromatic MAGPHYS fit yields $M_d = 5.7_{-0.1}^{+0.46} \times 10^6 M_\odot$. This quantifies the dust mass discrepancy at a factor between $7-18$. From the optical data, we only found dust inside the dust lane. The spatial resolution of the FIR/sub-mm data was not high enough to disentangle any potential dust at higher scale heights.

\citet{Wise1996} argue that a diffusely distributed dust component, invisible in single-band optical images, can be traced by means of broadband optical colour gradients. Using a diffuse dust model, they show that the gradients expected from \textit{IRAS}-based dust masses are compatible with the observed gradients. More precisely, if the observed colour gradients were completely due to dust attenuation, the resulting dust masses would be significantly higher than the \textit{IRAS} dust masses. They note that age and metallicity gradients may also contribute. Therefore, their dust masses may be considered as upper limits. In the case of NGC 4370, we found that the colour gradients outside the dust lane are relatively flat, suggesting only minor dust reddening.

On the other hand, a diffuse dust component is somewhat at odds with more recently derived colour gradients \citep{Michard2005}; models with significant amounts of diffuse dust have colour gradients that contradict the observed ones. In fact, such objects would completely destroy the correlation they find between B-R and B-V colours. Additionally, \citet{Vanderbeke2011} find a one-to-one correspondence in stellar velocity dispersions in the optical and near-infrared. Comparing this to theoretical models \citep{Baes2000,Baes2002} suggests there is little effect of diffuse dust on the stellar kinematics. They conclude that ETGs are virtually optically thin.

Thanks to the new possibilities offered by FitSKIRT, we are able to simulate the dust-starlight interaction properly. The automated inverse RT fitting makes it  possible to determine the best parameters in an unbiased way. The only assumption that is being made is the choice of dust geometry. We fit two axisymmetric dust geometries to mimic the dust distribution in the dust lane: an exponential disc and a ring. We do not add a second, diffuse component to the model. The exponential disc geometry yields a total dust mass of $ (5.0 \pm 2.4) \times 10^6\, M_\odot$, and the ring geometry yields $ (3.3 \pm 2.7) \times 10^6\, M_\odot$. While the disc produces slightly higher dust masses than the ring, both values are close and fall within their respective error bars. The uncertainties on these masses are in fact the rms of the dust masses of five equivalent simulations, run for each geometry. We found that the $\chi^2$ values for each of the simulations lie very close together, but the resulting dust masses vary significantly, hence the relatively high rms values. Most simulations actually yield a dust mass that is higher than the one with the lowest $\chi^2$, so the uncertainties on the dust masses should be seen as possible fluctuations towards the high end. With this in mind, we can say that the total dust masses from the RT fits are fully consistent with the FIR/sub-mm and panchromatic SED fit estimates.

Additionally, we tested an alternative stellar distribution derived from the IRAC 3.6 $\mu$m image to minimize the effects of dust obscuration and include the nuclear stellar disc. We only explore a ring geometry. The global model quality from this set-up is lower than the previous set-ups. The main issue seems to be the resolution, which causes a slightly different inclination angle and a stretched-out stellar profile. The residual features at the location of the dust lane are similar to the ones from the previous methods, but are generally larger. We find a total dust mass of $ (7.0 \pm 1.8) \times 10^6\, M_\odot$, which is slightly larger, but consistent with the other RT methods. In this manner, it seems that the effect of a nuclear disc on the dust mass determination is minor. The dust mass from this model is also fully consistent with the FIR/sub-mm and panchromatic SED fit estimates.  However, given the strong residual features, we do not prefer this model to the other RT fits.

We must point out that the dust mass derived from FitSKIRT may be overestimated. An exponential disk has its highest density in the centre, which is unlikely for the case of NGC 4370. Second, the density in this geometry only falls to zero at infinite radii. Dust extinction is visible in the outskirts of the galaxy in the disc model, while this does not seem to be the case in the observed images. These two arguments may lead to an overestimation of the dust mass. The ring geometry does not suffer from the above problems. Inherent to the analytical description of the geometry, there is less dust in the centre and almost none at large radii. One can therefore assume that the representation of the dust geometry by a ring is better. A final argument, applicable to both geometries, is a line-of-sight effect. NGC 4370 is an edge-on galaxy, meaning we only see one side of the dust lane. It is possible that the dust distribution does not extend to the other side of the galaxy. These so-called "arcs" or incomplete dust structures have been observed in several dusty ETGs with different inclinations \citep{Goudfrooij1994b,Patil2007,Finkelman2008}. In the extreme case that the other side of the galaxy is completely free of dust, this can cause an overestimation of the total dust mass by a factor of 2. This underlines the importance of a realistic dust geometry and the proper treatment of absorption and multiple, anisotropic scattering to derive dust masses from optical data. 

Our models do not assume anything about the emission of the dust in NGC 4370. Such energy balance studies have been conducted for several spiral galaxies \citep[e.g.][]{Baes2010,Popescu2011,DeLooze2012b,DeGeyter2015} and for the Sombrero galaxy \citep[M104, ][]{DeLooze2012a}. However, this requires assumptions on the dust emissivity, sub-resolution physics (clumps and filaments) and a thorough knowledge of the dust-heating mechanisms: old and young stars, cosmic rays, and potentially  external X-ray heating (NGC 4370 lies close to massive ellipticals in the Virgo cluster). All this requires further investigation and falls beyond the scope of this paper.

Foreground screen and uniform mix models run short to represent the 3D intrinsic dust geometry. They should be seen as lower limits. Specifically in NGC 4370, RT simulations show that the dust is modelled best with an exponential disc or ring geometry. The latter produces lower residuals and images that are more similar to the observations. The total mass of the dust in these geometries is slightly lower, but consistent with \textit{Herschel}-derived dust masses within 1 rms. Our results suggest that it is premature to invoke a diffuse dust component to explain the amount of dust that is observed in NGC 4370.

\section{Conclusions} \label{sec:conclusions}

In this paper, we explored several methods of deriving the total dust mass in the dust-lane ETG NGC 4370. Table \ref{tab:dustmasses} gives an overview of the total dust mass for all methods. We briefly summarize our findings here.

- Exploiting the sensitivity and wavelength range of \textit{Herschel}, we were able to constrain the dust mass very well within the limits of the dust model. We find a dust mass that is an order of magnitude higher than previous estimates derived from IRAS fluxes.

- We constructed colour maps based on high-quality $ugiz$ data and found a lower limit to the dust mass, which lies an order of magnitude below the FIR/sub-mm-based estimate.

- We find a dust mass comparable to the colour-derived mass through MGE fits to the stellar distribution of NGC 4370. These dust-free MGE models allow the creation of attenuation maps when compared to observations. Assuming a simple foreground screen geometry for the dust, this yields dust masses comparable to the average colour-derived mass. For a uniform mixing of dust and stars, we found dust masses that are two to three times higher.

- Since neither of these relatively simple methods yields a dust mass that is high enough to match the FIR/sub-mm dust mass, we resorted to the more advanced method of radiative transfer simulations. Using FitSKIRT, an inverse radiative transfer optimizer around the radiative transfer code SKIRT, we could evaluate more realistic dust geometries: an exponential disc and a ring. Both geometries resulted in a decent fit, however the ring geometry is preferred. The dust masses of both models are slightly lower, but fully consistent with the FIR/sub-mm dust masses within 1 rms.

- We checked the influence of a nuclear stellar disc by adopting the stellar distribution from the IRAC 3.6 $\mu$m image instead of the $z$ band image. We only explored the ring geometry, but found a best fit model of lower quality than the previous FitSKIRT models. The dust mass is a fraction higher, but consistent with these models and with the FIR/sub-mm dust masses. This suggests that the influence of a nuclear stellar disc is limited when determining the dust mass in NGC 4370.

- Within the current models, there is no need to invoke a diffuse dust component for NGC 4370. Instead, we suspect that radiative transfer simulations can reproduce the true dust mass with only dust in the dust lane, provided an accurate approximation of the intrinsic dust geometry is made.

These findings do not imply that the more empirical methods (colour maps and MGE fits) in itself are incorrect. If one adopted a proper description of a ring geometry and constructed the attenuation - optical depth formula as in equations \eqref{eq:screen} or \eqref{eq:sandwich}, these methods would yield more reliable dust masses. We therefore caution against over-interpreting dust masses and optical depths based on optical data alone, when using overly simplistic star-dust geometries. In fact, for NGC 4370, the difference in total dust mass between the simple screen geometry and the 3D radiative transfer fits is a factor of $\sim$10. 

We aim to expand this effort to a set of ten early-type galaxies with prominent dust lanes. The objects in this sample have all been observed with \textit{Herschel} as part of the Far-infraRed Investigation of Early-type galaxies with Dust Lanes (FRIEDL). This will allow us to check the conclusions of this work against similar objects.

\begin{acknowledgements}

S.V, G.D.G., and M.B. gratefully acknowledge the support of the Flemish Fund for Scientific Research (FWO-Vlaanderen). IDL is a postdoctoral researcher of the FWO-Vlaanderen (Belgium). M.B. and J.F. acknowledge financial support from the Belgian Science Policy Office (BELSPO) through the PRODEX project "\textit{Herschel}-PACS Guaranteed Time and Open Time Programs: Science Exploitation" (C90370). \\
The authors wish to thank Anthony Jones for an insightful discussion of dust physics and SED fitting. \\
This work has been realized within the CHARM framework (Contemporary physical challenges in Heliospheric and AstRophysical Models), a phase VII Interuniversity Attraction Pole (IAP) programme organised by BELSPO, the BELgian federal Science Policy Office. \\
This work is based on observations obtained with MegaPrime/MegaCam, a joint project of CFHT and CEA/DAPNIA, at the Canada–France–Hawaii Telescope (CFHT), which is operated by the National
Research Council (NRC) of Canada, the Institut National des Sciences de l'Univers of the Centre National de la Recherche Scientifique (CNRS) of France and the University of Hawaii. This work has been supported in part by the Canadian Advanced Network for Astronomical Research (CANFAR) through funding from CANARIE under the Network-Enabled Platforms programme. This research made use of the facilities at the Canadian Astronomy Data Centre, which are operated by the National Research Council of Canada with support from the Canadian Space Agency. \\
We thank all the people involved in the construction and the launch of \textit{Herschel}. SPIRE was developed by a consortium of institutes led by Cardiff University (UK) and including Univ. Lethbridge (Canada); NAOC (China); CEA, LAM (France); IFSI, Univ. Padua (Italy); IAC (Spain); Stockholm Observatory (Sweden); Imperial College London, RAL, UCL-MSSL, UKATC, Univ. Sussex (UK); and Caltech, JPL, NHSC, Univ. Colorado (USA). This development has been supported by national funding agencies: CSA (Canada); NAOC (China); CEA, CNES, CNRS (France); ASI (Italy); MCINN (Spain); SNSB (Sweden); STFC and UKSA (UK); and NASA (USA). HIPE is a joint development (are joint developments) by the \textit{Herschel} Science Ground Segment Consortium, consisting of ESA, the NASA \textit{Herschel} Science Center, and the HIFI, PACS and SPIRE consortia.\\
\end{acknowledgements}

\bibliographystyle{aa} 
\bibliography{allreferences}

\end{document}